\documentclass[trackchanges,twocolumn]{aastex7}
\usepackage{amsmath}
\usepackage{longtable}
\usepackage{changepage}
\usepackage{rotating}
\usepackage{silence}
\usepackage{graphicx}
\usepackage{subcaption}
\usepackage{float}
\usepackage{xcolor}

\usepackage{hyperref}

\providecommand{\sw}[1]{\texttt{#1}}

\newcommand{\thisgrb}{GRB~250704B}
\newcommand{\jetsimpy}{\sw{jetsimpy}}
\newcommand{\afterglowpy}{\sw{afterglowpy}}
\newcommand{\multi}{\sw{multi-nest}}

\newcommand{\giteam}{}

\begin{document}

\title{\thisgrb: An Off-axis Short GRB with a Long-Lived Afterglow Plateau}

\author[0000-0002-7942-8477]{Vishwajeet Swain}
\altaffiliation{These authors contributed equally to the manuscript.}
\email{vishwajeet.s@iitb.ac.in}
\affiliation{Department of Physics, Indian Institute of Technology Bombay, Powai, Mumbai 400076, India}

\author[0000-0002-2184-6430]{Tomás Ahumada}
\altaffiliation{These authors contributed equally to the manuscript.}
\email{tahumada@astro.umd.edu}
\affiliation{Cahill Center for Astrophysics, California Institute of Technology, MC 249-17, 1216 E California Boulevard, Pasadena, CA, 91125, USA}

\author[0009-0002-8110-0515]{Sameer K. Patil}
\email{sameerkpatil@gmail.com}
\affiliation{Department of Physics, Indian Institute of Technology Bombay, Powai, Mumbai 400076, India}

\author[0000-0002-5890-9298]{Yogesh Wagh}
\email{ywagh4430@gmail.com}
\affiliation{Department of Physics, Indian Institute of Technology Bombay, Powai, Mumbai 400076, India}

\author[0000-0002-6112-7609]{Varun Bhalerao}
\email{varunb@iitb.ac.in}
\affiliation{Department of Physics, Indian Institute of Technology Bombay, Powai, Mumbai 400076, India}

\author[0000-0002-4534-7089]{Ehud Nakar}
\email{ehud.nakar@gmail.com}
\affiliation{School of Physics and Astronomy, Tel Aviv University, Tel Aviv 6997801, Israel}

\author[0000-0002-5619-4938]{Mansi Kasliwal}
\email{mansi@astro.caltech.edu}
\affiliation{Cahill Center for Astrophysics, California Institute of Technology, MC 249-17, 1216 E California Boulevard, Pasadena, CA, 91125, USA}

\author[0000-0002-9364-5419]{Xander J. Hall}
\email{xjh@andrew.cmu.edu}
\affiliation{McWilliams Center for Cosmology and Astrophysics, Department of Physics, Carnegie Mellon University, 5000 Forbes Avenue, Pittsburgh, PA 15213}

\author[0009-0001-0574-2332]{Malte Busmann}
\email{}
\affiliation{University Observatory, Faculty of Physics, Ludwig-Maximilians-Universität München, Scheinerstr. 1, 81679 Munich, Germany}

\author[0000-0003-3768-7515]{Shreya Anand}
\altaffiliation{LSST-DA Catalyst Postdoctoral Fellow}
\email{sanand08@stanford.edu}
\affiliation{Kavli Institute for Particle Astrophysics and Cosmology, Stanford University, 452 Lomita Mall, Stanford, CA 94305, USA}
\affiliation{Department of Astronomy, University of California, Berkeley, CA 94720-3411, USA}

\author[0000-0003-2758-159X]{Viraj Karambelkar}
\email{}
\affiliation{Cahill Center for Astrophysics, California Institute of Technology, MC 249-17, 1216 E California Boulevard, Pasadena, CA, 91125, USA}


\author[0000-0002-8977-1498]{Igor Andreoni}
\email{igor.andreoni@unc.edu}
\affiliation{Department of Physics and Astronomy, University of North Carolina at Chapel Hill, Chapel Hill, NC 27599, USA}

\author[0000-0003-3533-7183]{G. C. Anupama}
\email{gca@iiap.res.in}
\affiliation{Indian Institute of Astrophysics, II Block Koramangala, Bengaluru 560034, India}

\author[0009-0007-9244-191X]{Anuraag Arya}
\email{arya.a@iitb.ac.in}
\affiliation{Department of Physics, Indian Institute of Technology Bombay, Powai, Mumbai 400076, India}

\author[0000-0003-0477-7645]{Arvind Balasubramanian}
\email{arvind.balasubramanian@iiap.res.in}
\affiliation{Indian Institute of Astrophysics, II Block Koramangala, Bengaluru 560034, India}

\author[0000-0002-3927-5402]{Sudhanshu Barway}
\email{sudhanshu.barway@iiap.res.in}
\affiliation{Indian Institute of Astrophysics, II Block Koramangala, Bengaluru 560034, India}

\author[0000-0001-8544-584X]{Jonathan Carney}
\email{jcarney@unc.edu}
\affiliation{Department of Physics and Astronomy, 
    University of North Carolina at Chapel Hill, 
    Chapel Hill, NC 27599-3255, USA}

\author[0000-0002-8262-2924]{Michael Coughlin}
\email{cough052@umn.edu}
\affiliation{School of Physics and Astronomy, University of Minnesota, Minneapolis, MN 55455, USA}

\author[0000-0001-7841-0294]{Deepak Eappachen}
\email{deepak.eappachen@iiap.res.in}
\affiliation{Indian Institute of Astrophysics, II Block Koramangala, Bengaluru 560034, India}

\author[0009-0006-7990-0547]{James Freeburn}
\email{jfreebur@unc.edu}
\affiliation{Department of Physics and Astronomy, University of North Carolina at Chapel Hill, Chapel Hill, NC 27599, USA}

\author[0000-0003-3270-7644]{Daniel Gruen}
\email{}
\affiliation{University Observatory, Faculty of Physics, Ludwig-Maximilians-Universität München, Scheinerstr. 1, 81679 Munich, Germany}

\author[0009-0001-4683-388X]{Tanishk Mohan}
\email{tanishk.mohan@iitb.ac.in}
\affiliation{Department of Physics, Indian Institute of Technology Bombay, Powai, Mumbai 400076, India}

\author[0000-0002-9700-0036]{Brendan O'Connor}
\email{}
\affiliation{McWilliams Center for Cosmology and Astrophysics, Department of Physics, Carnegie Mellon University, Pittsburgh, PA 15213, USA}

\author[0000-0002-6011-0530]{Antonella Palmese}
\email{apalmese@andrew.cmu.edu}
\affiliation{McWilliams Center for Cosmology and Astrophysics, Department of Physics, Carnegie Mellon University, Pittsburgh, PA 15213, USA}

\author[0009-0002-7897-6110]{Utkarsh Pathak}
\email{utkarshpathak.07@gmail.com}
\affiliation{Department of Physics, Indian Institute of Technology Bombay, Powai, Mumbai 400076, India}

\author[0000-0002-6688-0800]{D. K. Sahu}
\email{dks@iiap.res.in}
\affiliation{Indian Institute of Astrophysics, II Block Koramangala, Bengaluru 560034, India}

\author[0009-0005-2987-0688]{Aditya Pawan Saikia}
\email{adityaps@iitb.ac.in}
\affiliation{Department of Physics, Indian Institute of Technology Bombay, Powai, Mumbai 400076, India}

\author[0000-0003-2700-1030]{Nikhil Sarin}
\email{nikhil.sarin@ast.cam.ac.uk}
\affiliation{Kavli Institute for Cosmology, University of Cambridge, Madingley Road, CB3 0HA, UK}
\affiliation{Institute of Astronomy, University of Cambridge, Madingley Road, CB3 0HA, UK}

\author[0000-0002-6428-2700]{Gokul Srinivasaragavan}
\email{gsriniv2@umd.edu}
\affiliation{Department of Astronomy, University of Maryland, College Park, MD 20742, USA}
\affiliation{Joint Space-Science Institute, University of Maryland, College Park, MD 20742, USA} 
\affiliation{Astrophysics Science Division, NASA Goddard Space Flight Center, 8800 Greenbelt Rd, Greenbelt, MD 20771, USA}

\author[0009-0008-6644-5412]{Hitesh Tanenia}
\email{hitesh.tanenia@iitb.ac.in}
\affiliation{Department of Physics, Indian Institute of Technology Bombay, Powai, Mumbai 400076, India}



\begin{abstract}
We present a detailed multi-wavelength afterglow study of the short \thisgrb, extensively monitored in optical and near-infrared bands. Its afterglow displays an unusually long-duration plateau followed by an achromatic break and a steep decline, deviating from canonical GRB afterglows. While long plateaus are often explained by central engine activity, we find that for \thisgrb, an energy injection model requires unreasonable parameters. The afterglow is better explained by an off-axis power-law structured jet with a narrow core ($\theta_c \approx 0.7^{\circ}$) viewed at a modest angle ($\theta_v \approx 1.9^{\circ}$). A comparison with GRB~170817A shows that both events are consistent with the off-axis structured jet scenario, where the shape of the light curve is governed primarily by the geometry of the jet and the viewing angle rather than the energetics, microphysical parameters, or external density. Our results underscore the importance of incorporating the jet structure in GRB modeling.
\end{abstract}



\section{Introduction}\label{sec:introduction}

Gamma-ray bursts (GRBs) are among the most luminous explosions in the Universe, characterized by an intense prompt $\gamma$-ray flash followed by a broadband afterglow \citep{meszaros2006}. The current classification of GRBs is based on the duration of their prompt emission. Bursts whose 90\% prompt emission is released in $T_{90}< 2$~s are classified as short GRBs, whereas bursts lasting longer than 2 s are considered long GRBs \citep{1993ApJ...413L.101K}. Some short GRBs exhibit extended emission in the $\gamma$-ray band after the initial short flash \citep[e.g.,][]{2002ARA&A..40..137M, 2006ApJ...643..266N, 2011AAS...21710803N}. Given the classical definition, short GRBs with extended emission (EE) can have a significantly longer $T_{90}$. Because EE is also spectrally softer, classification based solely on $T_{90}$ remains debated \citep{ahumada2021grb,zhang2021,rastinejad2022,troja2022,Levan2024grb,grb211211yang}.

Gamma-ray burst afterglows, produced by the deceleration of relativistic ejecta in the circum-burst medium, are typically modeled as synchrotron emission from a forward shock. Their light curves often follow a power-law decay in time, punctuated by breaks that can arise from changes in the dynamics or geometry of the outflow. A common cause for such breaks is a “jet break," when the relativistic beaming angle exceeds the physical opening angle of the jet, leading to a faster decline in flux \citep{sari1999,Granot_2002,oConnor2024xrayjets}. While on-axis afterglows display the canonical bright-to-faint behavior, off-axis afterglows rise more slowly and peak later, as the relativistic beaming cone gradually widens into the observer's line of sight; GW170817/GRB 170817A is a prime example of such off-axis geometry \citep{2018ApJ...867...57R, 2019MNRAS.489.1919T, 2020ApJ...896..166R, 2021ApJ...922..154M}. 
By modeling afterglow light curves across wavelengths and incorporating parameters such as jet geometry and observer viewing angle, one can disentangle typical, off-axis, and dark afterglow behaviors and explain the presence and timing of breaks in their evolution.

Some GRB afterglows display an early-time plateau phase, where the light curve remains nearly flat before transitioning into the standard power-law decay. In a purely geometric framework, plateaus can occur in off-axis events when the observer's line of sight is just outside the jet core: as the relativistic beaming cone gradually widens, the observed flux increases or stays constant before declining, producing a plateau-like feature. This effect is much more common in X-ray afterglows, where more than half of Swift-detected bursts show plateaus, whereas in the optical they are relatively rare—only a few dozen have been reported (e.g., GRB 120404A, GRB 140903A, GRB 150424A, GRB 231117A) \citep{grb120404A,grb140903A,grb150424A,grb231117A}. Other explanations for plateaus, such as sustained central engine activity from a magnetar or late-time energy injection into the blast wave, have been proposed \citep{2001ApJ...552L..35Z, 2008MNRAS.385.1455M, Rowlinson_2013}.

Population studies increasingly favor angularly structured jets over simple top-hat jets for both long and short GRBs: modeling shows that a narrow core ($\approx 3 - 5$ deg) with shallower wings can reproduce observed afterglow diversity, luminosity functions, and event rates. The clearest case is GRB 170817A/GW170817, where late-time radio/X-ray evolution and VLBI superluminal motion require a successful, narrowly collimated core embedded in wider-angle ejecta \citep{2018Natur.561..355M,Mooley2022}. Beyond 170817A, several bursts show afterglow behavior best explained with structure or modest off-axis viewing, including GRB 150101B (a 170817A-like analog at cosmological distance), GRB 160821B (afterglow+kilonova modeling probes jet geometry), and, among long GRBs, the extreme GRB 221009A, whose broadband afterglow prefers a shallow structured jet \citep{GRB150101B, GRB160821Bmagic,GRB160821Blamb,GRB160821Btroja,GRB221009Agrandma,gokul221009A,grb221009Aoconnor}. Recent catalog-level analyses of short GRBs further use afterglow light-curve shapes and viewing-angle constraints to argue that structured jets may be common rather than exceptional.

The short \thisgrb\ displayed an unusual afterglow: a one-day plateau followed by a sharp achromatic break and rapid decay, distinguishing it from the known short GRB population. We present results from extensive multi-wavelength follow-up and broadband modeling. In \S\ref{sec:observations}, we describe our observations and data reduction, covering X-ray to radio bands, and list the publicly available datasets used in this work. In \S\ref{sec:prompt}, we briefly summarize the prompt properties of this GRB, based on public data; detailed prompt-emission analysis is beyond the scope of this paper. Section~\ref{sec:afterglow-analysis} discusses the temporal and spectral behavior of the afterglow. In \S\ref{sec:afterglow_model}, we present broadband modeling of the afterglow. Finally, in \S\ref{sec:discussion}, we summarize our results and compare this GRB with the GW170817 counterpart.

\section{Observations and Data}\label{sec:observations}

\giteam \thisgrb\ was first reported by The Space Variable Objects Monitor -- Gamma Ray burst Monitor (\emph{SVOM}-GRM) with trigger time $T_0$= 2025-07-04T08:16:27 UT \citep{2025GCN.40940....1S}. The prompt emission shows a short burst consisting of two episodes with $T_{90} = 0.68 \pm 0.15$~s in the $15-5000$~keV band. Several other instruments also reported this burst including the Einstein Probe -- Wide Field Telescope \citep[\emph{EP}-WXT;][]{2025GCN.40956....1L}, \emph{Konus}-Wind \citep{2025GCN.40972....1F}, \emph{Insight}--Hard X-ray Modulation Telescope \cite[HXMT;][]{2025GCN.40978....1W}, and the \emph{CALET} -- Gamma-Ray Burst Monitor \citep{2025GCN.41025....1S}. The Inter-Planetary Network (\emph{IPN}) also reported the detection and triangulation of this burst \citep{2025GCN.41050....1K}.

\giteam An optical counterpart of \thisgrb\ was first reported by the COLIBRI at position RA (J2000): $20^h 03^m 29.51^s$ and Dec (J2000): $13\degr 01\arcmin 23.46\arcsec$, with an uncertainty of 0.5\arcsec\ \citep{2025GCN.40942....1S}. Independent VLT/FORS2 observations obtained a redshift of 0.661 \citep{2025GCN.40945....1M}, which we utilize throughout this paper. For broad-band follow-up observations, we triggered a number of telescopes: GROWTH-India Telescope (GIT), Himalayan Chandra Telescope (HCT), W. M. Keck Telescope (Keck), Victor M. Blanco Telescope (Blanco), Fraunhofer Telescope at Wendelstein Observatory (FTW), Southern Astrophysical Research Telescope (SOAR), Palomar 200-inch, Palomar 60-inch, Giant Metrewave Radio Telescope (GMRT), as part of the GROWTH Collaboration \citep{2019PASP..131c8003K}. We also used data from various circulars reported on the General Coordinate Network (GCN). All data and their sources are listed in Table~\ref{tab:afterglow_obs_table}. The observations and data reduction are described in the Appendix \ref{appendixData}. 

\section{Prompt emission: analysis}\label{sec:prompt}
\giteam The prompt emission of the short \thisgrb\ was detected by multiple satellites. In this work, we adopt the results reported by \emph{Konus}-Wind \citep{2025GCN.40972....1F}. The observed light curve consists of two distinct episodes, which can be interpreted as phases of central engine activity separated by a brief quiescent interval of $\sim 0.1$~s. Each episode contains multiple distinct pulses, and the total burst duration is $T_{90} \approx 0.4$s. They report a measured total fluence of $(4.24 \pm 0.65) \times 10^{-6}\ \mathrm{erg\ cm^{-2}}$ in the 20~keV -- 10~MeV energy range. The 16~ms peak flux, measured from $T_0 + 0.240$~s, is $(5.82 \pm 0.89) \times 10^{-5}\ \mathrm{erg\ cm^{-2}\ s^{-1}}$. The time-integrated spectrum from $T_0$ to $T_0 + 0.256$ s is best fit with a Band function in the 20~keV -- 15~MeV range, with parameters $\alpha = -1.17^{+0.9}_{-0.8}$, $\beta = -2.48^{+0.39}_{-1.91}$, and an observed peak energy of $E_{p, obs} = 935^{+305}_{-197}$~keV.

\giteam Using a redshift of $z = 0.661$, they calculated the isotropic-equivalent energy of the burst to be $E_{\mathrm{iso}} = (5.15 \pm 0.79) \times 10^{51}$~erg, and the rest-frame peak energy is $E_{p} = 1550^{+510}_{-330}$~keV. Overall, \thisgrb\ exhibits a hard spectrum, consistent with the typical characteristics of short GRBs in their prompt emission. 

\giteam \emph{EP}-WXT detected the transient in the soft energy range of ($0.5-4$)~keV, starting at the same $T_0$ and lasting for 10~s before the observation was interrupted by the autonomous follow-up observation \citep{2025GCN.40956....1L}. According to the \emph{EP}-WXT report, the averaged unabsorbed flux is $1.3 \pm 0.95 \times 10^{-9}\ \mathrm{erg/cm^2/s}$ and the corresponding photon index is $1.7 \pm 1.3$ over the pulse of 10~s. We note that since the observation was terminated by the slew, we cannot comment on whether this might be an extended tail of the prompt emission, or the detection of the early afterglow.

\section{Afterglow}\label{sec:afterglow-analysis}

\giteam The interaction of the GRB jet with the circum-burst medium produces synchrotron radiation, observed as a multi-wavelength afterglow that probes the burst energetics and environment \citep{Sari_1997, Granot_2002}. In the simplest framework, the afterglow emission can be described by simple power-law dependencies in both time and frequency, expressed as $F_{\nu} \propto t^{-\alpha} \nu^{-\beta}$. All afterglow data used in this work are given in Table~\ref{tab:afterglow_obs_table}, \ref{tab:xray-data}, and \ref{tab:radio-data}.

\subsection{X-ray afterglow}\label{sec:afterglow_xray}
\begin{figure}[ht]
        \centering
    \includegraphics[width=\columnwidth]{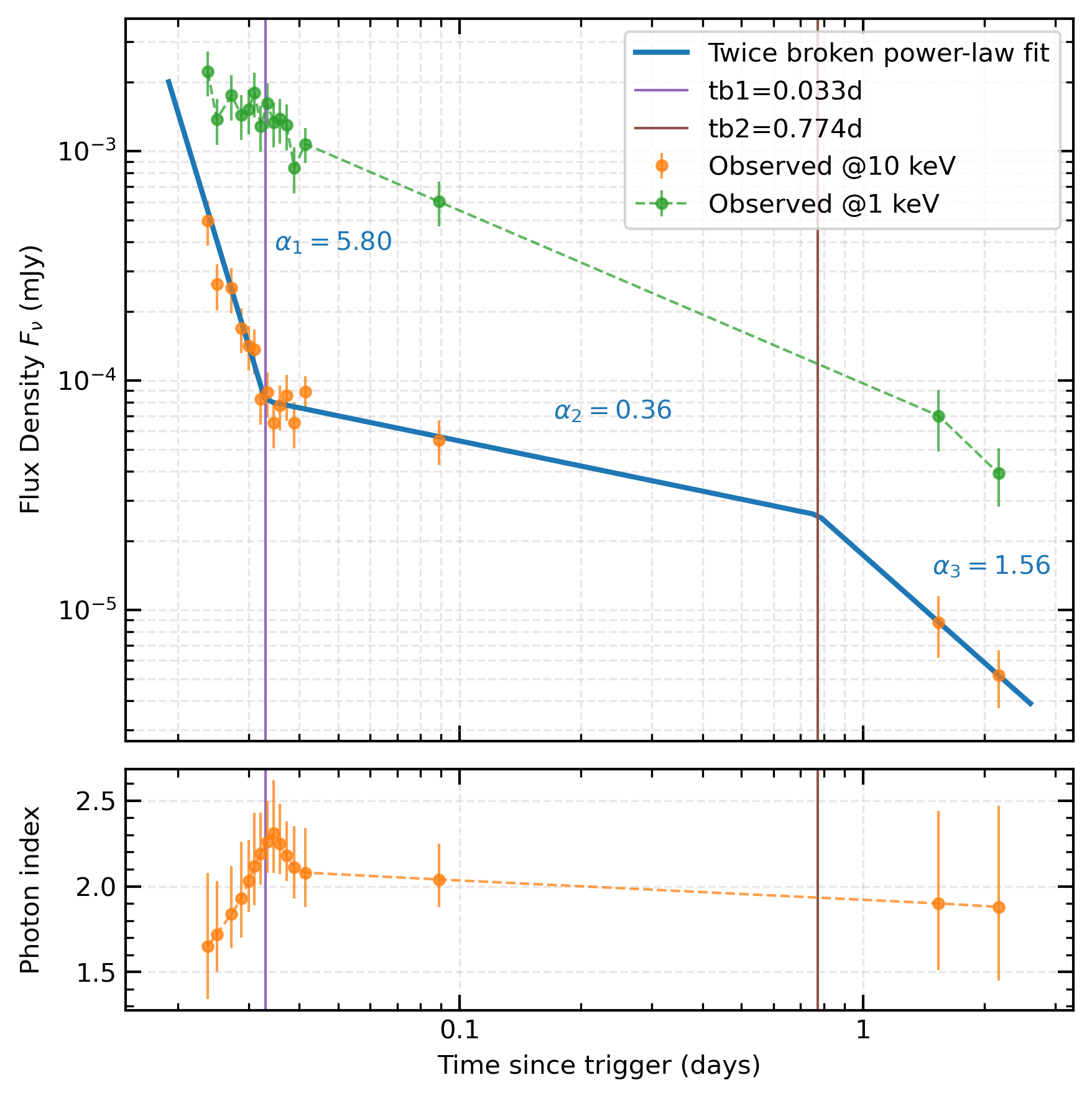}
    \caption{The upper panel shows flux densities calculated at 1~keV (green) and 10~keV (orange). The 10~keV lightcurve is fit with a twice-broken power-law fit (blue). The two temporal break are identified at $t_{b1} = 0.033$~d (purple vertical line) and $t_{b2} = 0.774$~d (brown vertical line), with decay indices $\alpha_1 = 5.8$, $\alpha_2 = 0.36$, and $\alpha_3 = 1.56$ marked along the fit. The bottom panel shows the evolution of the photon index. Prior to the first break, the 10 keV flux density exhibits an excess emission with rapid decay, while the 1 keV flux density remains nearly flat.} 
    \label{fig:swift-xrt}
\end{figure}

\giteam Figure~\ref{fig:swift-xrt}, shows the \emph{Swift}-XRT light curve at 10~keV and 1~keV. The 10~keV flux density exhibits steep initial decay, followed by a much shallower decline. A broken power-law fit yields an initial slope of $\alpha_{X1} = 5.8$, a post-break slope of $\alpha_{X2} = 0.36$, and a break time of $t_{b,X} = 0.03$~d. The steep early decline may be attributed to high-latitude emission \citep{2020A&A...641A..61A}, although we do not explore this interpretation further here. In contrast, the 1~keV flux density does not show such a rapid decay. This difference arises from spectral evolution: as the spectrum evolves from hard to soft, the 10~keV flux density decays more steeply in the early phases. To avoid contamination from this component, we exclude XRT data prior to $t_{b,X}$ from subsequent analysis.

\begin{figure*}[ht]
    \centering
    \includegraphics[width=\textwidth]{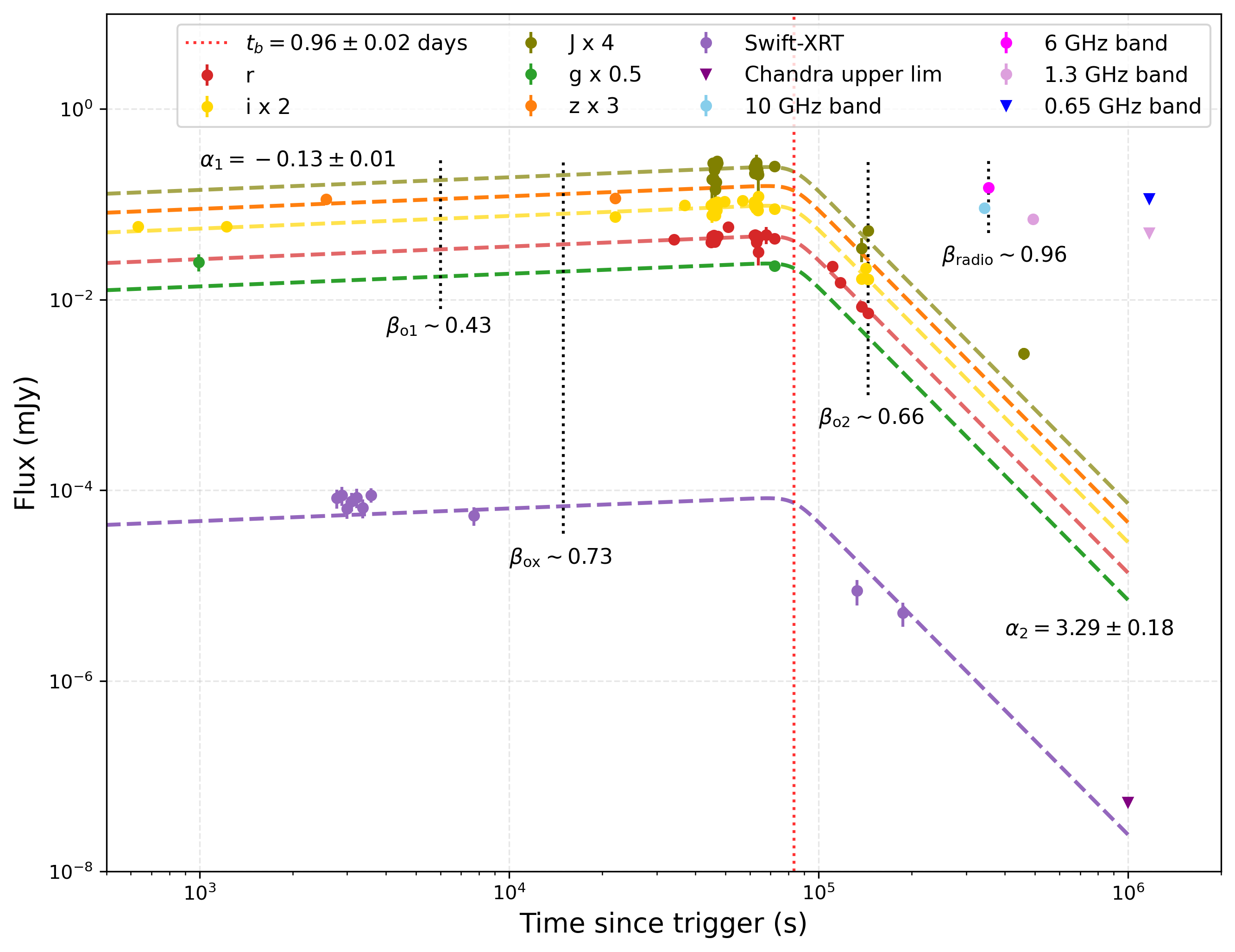}
    \caption{Multi-wavelength afterglow light curves of \thisgrb\ in X-ray (purple), optical ($r$, $i$, $z$, $g$, and $J$ bands), and radio (1.3, 6, and 10~GHz, with an upper limit at 0.65~GHz). The light curves are well described by a broken power-law with an initial shallow plateau phase ($\alpha_1 = -0.13 \pm 0.01$) followed by a steep decay ($\alpha_2 = 3.29 \pm 0.18$) after the break at $t_b = 0.96 \pm 0.02$ days (red vertical line). Spectral indices from the fits are $\beta_{\mathrm{o1}} = 0.43 \pm 0.06$ (optical) and $\beta_{\mathrm{ox}} = 0.73 \pm 0.02$ (optical-to-X-ray), while those derived from observations are $\beta_{\mathrm{o2}} = 0.66 \pm 0.08$ (optical) and $\beta_{\mathrm{radio}} = 0.96 \pm 0.16$ (radio). The consistent temporal evolution across X-ray, optical, and radio bands indicates an achromatic break, supporting an off-axis jet interpretation.}
    \label{fig:light-curve}
\end{figure*}

\subsection{Temporal evolution}\label{sec:afterglow_temporal}
\giteam We first fit a simple model to the afterglow data to ascertain its basic properties. Our multi-wavelength data set spans X-ray to Radio bands. The dataset is particularly rich in the optical $r$, $i$, and the Infra-red $J$ bands. For X-ray analysis, we use 10~keV data from \emph{Swift}-XRT as discussed in \S\ref{sec:afterglow_xray}.

\giteam \thisgrb\ has a galactic latitude of $-10.04^\circ$, hence galactic extinction ($A_r = 0.3$) cannot be ignored. We corrected optical data for galactic extinction using \citet{2011ApJ...737..103S}, while $J$ band data were corrected using \citet{1998ApJ...500..525S}. Note that the unabsorbed X-ray fluxes are calculated using the best-fit $N_H$ values rather than just the galactic ones. The combined light curve exhibits an extended plateau lasting $\sim$1~day, followed by a rapid decay. We modeled it using a smooth broken power-law:
\begin{equation}
F(x) = A \, \left(\frac{t}{t_b}\right)^{-\alpha_1} 
\left[ \frac{1}{2} \left( 1 + \left(\frac{t}{t_b}\right)^{1/\delta} \right) \right]^{(\alpha_1 - \alpha_2)\,\delta},
\end{equation}
where A is the normalization, $t_b$ the break time, $a_1$ and $a_2$ are the temporal decay indices before and after the break, and $\delta$ controls the smoothness of the transition \citep{2013A&A...558A..33A}. Note that the X-ray fit discussed in \S\ref{sec:afterglow_xray} does not have this smoothing parameter. We assume an achromatic break to perform a joint fit to the $r$, $i$ and $J$ data to obtain $\alpha_1 = -0.13 \pm 0.01$, $\alpha_2 = 3.28 \pm 0.18$, $t_b = 0.96 \pm 0.02$~d, and $\delta = 0.06 \pm 0.05$: confirming a plateau followed by a steep decay. We then scale this achromatic power-law to the other bands, and show all results in Figure~\ref{fig:light-curve}. We find that the optical trend is reasonably followed in the X-ray band too --- however, our late-time $J$ band data point is inconsistent with this simplistic model. The achromatic nature of the break suggests that we may be seeing the evolution of a structured jet, or a jet break: though the latter is typically not preceded by a plateau \citep{2009ApJ...698.1261Z}. No additional breaks are seen in our light curve up to $\sim 1.67$~days.

\subsection{Spectral properties}\label{sec:afterglow_spectral}
\giteam Assuming a power-law spectrum $F_\nu \propto \nu^{-\beta}$, we now estimate $\beta$ at various points in the light curve. Our observations are not uniformly spaced, leaving some regions where we have coverage only in a certain band, or some spans with no coverage. Thus, we cannot directly measure the spectral slope at all points. Instead, we estimate fluxes in various bands from our broken power-law fit (\S\ref{sec:afterglow_temporal}) and use it to measure $\beta$. Since we have assumed the afterglow evolution including the break to be achromatic, we get $\beta_{\mathrm{o1}} = 0.43 \pm 0.06$. Using contemporaneous observations after the break at $(1.2-1.4)\times10^5$~s, we obtain $\beta_{\mathrm{o2}} = 0.66 \pm 0.08$ (Figure~\ref{fig:light-curve}).

\giteam Extending this approach to include X-rays, we calculated the optical-to-X-ray spectral index, giving $\beta_{\mathrm{ox}} = 0.73 \pm 0.02$. In the radio regime, detections were obtained at 6~GHz and 10~GHz at about $T_0 + 4$~days separated by just 2.88 hours. We ignore the small time separation and use these values to obtain $\beta_{\rm radio} = 0.96 \pm 0.16$, inconsistent with the optical and X-ray values. Overall, \thisgrb\ has a positive $\beta$ in the light curve in all observed bands.

\section{Afterglow modeling}\label{sec:afterglow_model}
\giteam In the standard fireball model of GRBs, the afterglow originates from synchrotron radiation produced when an ultra-relativistic jet interacts with the circum-burst medium (CSM) \citep{1992MNRAS.258P..41R,  1997ApJ...476..232M, 1998ApJ...497L..17S, Granot_2002}. The observed temporal and spectral evolution is sensitive to both the physical properties of the jet and the nature of the surrounding medium. By modeling this emission, one can infer key macro-physical parameters such as the isotropic equivalent kinetic energy ($E_\mathrm{K,iso}$), jet opening angle ($\theta_c$), and the observer viewing angle ($\theta_v$); as well as micro-physical parameters including the electron power-law index ($p$), the fraction of energy in relativistic electrons ($\epsilon_e$) and magnetic fields ($\epsilon_b$), and the fraction of accelerated particles ($\chi$). In this work, we assume $\chi = 1$, such that all accelerated electrons contribute to non-thermal synchrotron emission. The temporal and spectral evolution of the afterglow depends on the synchrotron break frequencies: the characteristic frequency ($\nu_m$), the cooling frequency ($\nu_c$), and the self-absorption frequency ($\nu_a$) \citep{Sari_1997, Granot_2002}. In addition, the viewing angle relative to the jet axis can strongly influence the observed light curve \citep{2002ApJ...570L..61G, 2022ApJ...940..189F}.

In this paper we explore an afterglow viewed off-axis with an structured jet, as it is the model that better describes the physics of this GRB. Our modeling for an on-axis jet with additional energy injection can be found in Appendix \ref{appendixEnergyInjection}.



\begin{figure}[ht]
    \centering
    \begin{subfigure}{\columnwidth}
        \includegraphics[width=\linewidth]{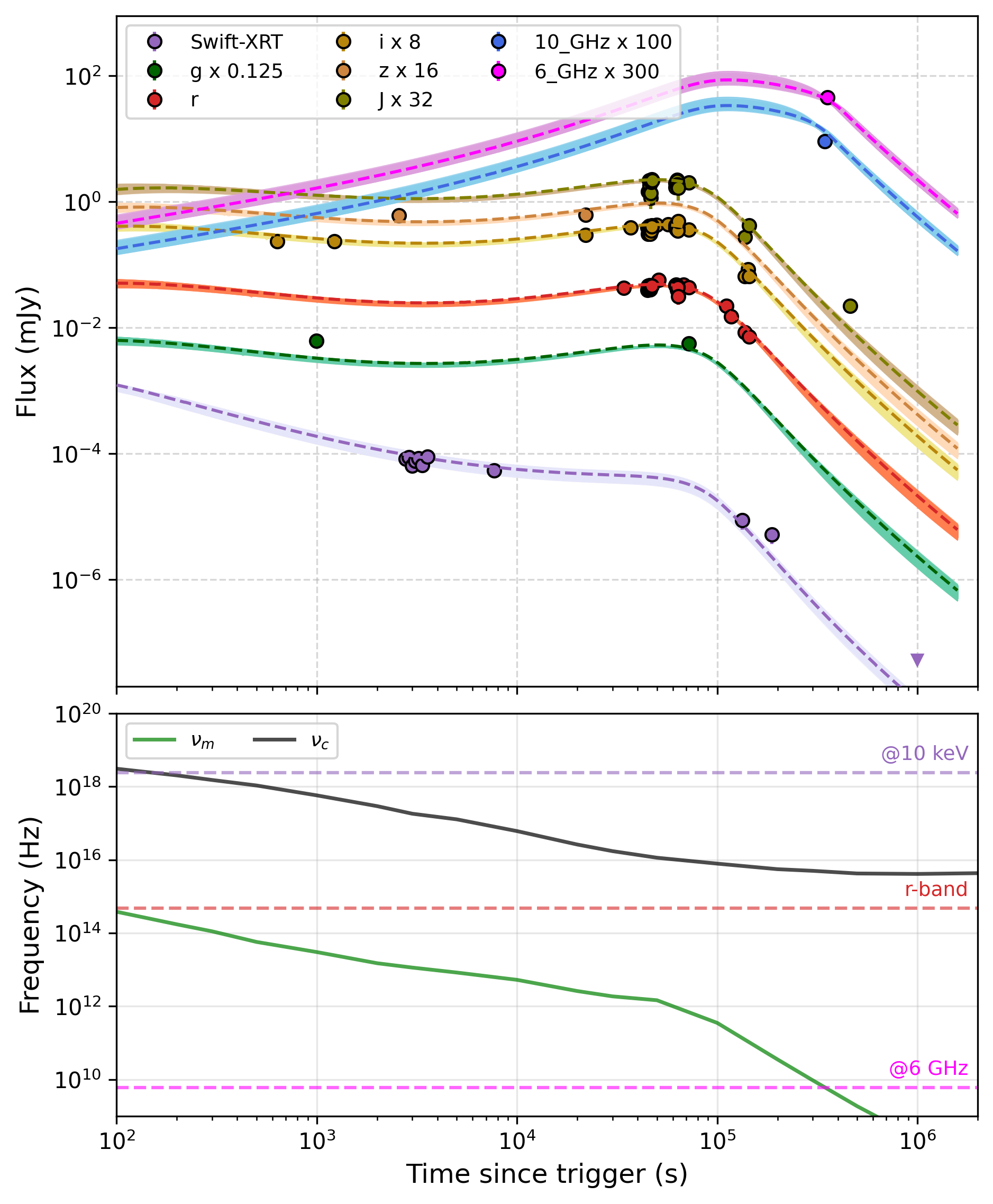}
        \caption{Upper panel: Multi-band afterglow light curves with dotted lines show the best-fit light curves, and shaded regions mark the $3\sigma$ uncertainties. Lower panel: Synchrotron break frequencies $\nu_m$ and $\nu_c$ as a function of time, calculated from model.}
        \label{fig:modelled_lc}
    \end{subfigure}
    
    \vspace{0.3cm}
    
    \begin{subfigure}{\columnwidth}
        \includegraphics[width=\linewidth]{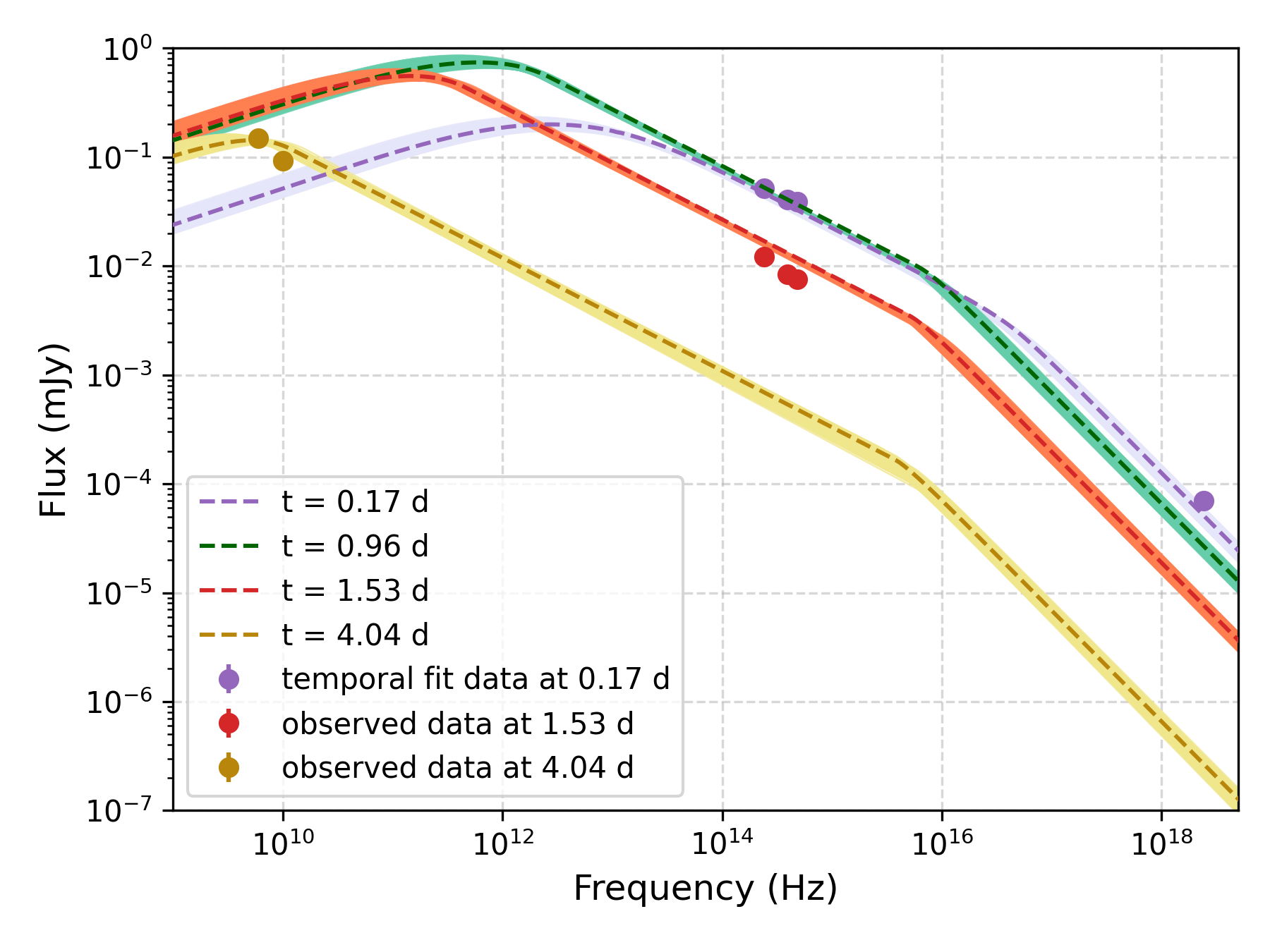}
        \caption{Spectral energy distributions at multi-epochs: 0.17~d (purple), t=0.96~d (green), t=1.53~d (red), and t=4.04~d (yellow) with corresponding derived spectrum from temporal-fit, observed optical–X-ray, and radio data .}
        \label{fig:model_spectra}
    \end{subfigure}
    
    \caption{Afterglow modeling of \thisgrb\ using \jetsimpy\, modeled with a power-law structured jet propagating into a uniform ISM viewed slightly off-axis.}
    \label{fig:afterglow_panels}
\end{figure}

\subsection{Data selection}\label{sec:datasel}
\giteam For afterglow modeling, we used all available optical data and the \emph{Swift}-XRT flux density at 10~keV. As discussed in \S\ref{sec:afterglow_xray}, we ignore initial XRT data. For both optical and X-ray data, synchrotron self-absorption is negligible because it affects only radio frequencies.

\giteam However, in the radio band, self-absorption plays a significant role at low frequencies. Thus, the flux evolution of the radio band depends not only on the cooling ($\nu_c$) and characteristic ($\nu_m$) frequencies but also on the absorption frequency ($\nu_a$). For short GRBs, the circum-burst environment is usually a uniform low-density interstellar medium (ISM) \citep{2020MNRAS.495.4782O, 2015ApJ...815..102F}. In this case, $\nu_a$ depends primarily on the ISM density ($n_0$) and remains constant during the afterglow. Low frequencies ($\sim$GHz) are often impacted by self-absorption, while higher frequencies like 6~GHz and 10~GHz are typically unaffected. Hence, we exclude the 1.3~GHz data from our modeling.

\subsection{Numerical modeling using \jetsimpy}\label{sec:jetsimpy}
\giteam We modeled the multi-wavelength afterglow using the publicly available package \jetsimpy\ \citep{2024ApJS..273...17W}, which calculates the synchrotron emission from a structured relativistic jet interacting with an external medium. The code adopts a reduced hydrodynamic model that approximates the blast wave as a thin 2-D surface, enabling efficient treatment of jet spreading at late times with reduced computational cost. 

\giteam We assumed a power-law jet defined as:
\begin{equation}
E(\theta) = E_{\mathrm{K,iso}} \left[ 1 + \left( \frac{\theta}{\theta_c} \right)^{2} \right]^{-s/2}, \label{eq:powerlaw_jet}
\end{equation}

\begin{equation}
\Gamma(\theta) = (\Gamma_{0} - 1) \left[ 1 + \left( \frac{\theta}{\theta_c} \right)^{2} \right]^{-s/2} + 1, \label{eq:powerlaw_jet2}
\end{equation}

where $E_{\mathrm{K, iso}}$ is the isotropic equivalent energy, $\Gamma_{0}$ is the initial Lorentz factor, $\theta_c$ is the half-opening angle of the jet and $s$ is the power law index \citep{2024ApJS..273...17W}. In our fits, we adopt a nominal value of $s=6$ \citep{2024ApJ...975..131R} and allow the jet to spread. The model does not incorporate synchrotron self-absorption. 

We initially allowed $\Gamma_{0}$ to vary as a free parameter, but the sampler consistently converged to very large values ($> 1000$). In contrast, fixing $\Gamma_{0}$ to a low value forced the model to compensate by requiring a very dense circum-burst medium, inconsistent with typical short GRB environments. Therefore, in our modeling, we have fixed $\Gamma_{0}$ to be sufficiently high ($10^{100}$) such that the blast wave begins directly in the deceleration phase, without an appreciable coasting stage. Note that jets with high Lorentz factor ($>1000$) have been seen in short GRBs, for example GRB~090510 \citep{2010ApJ...716.1178A}.

\giteam We then constrained the model parameters ($E_{\mathrm{K,iso}}$, $\epsilon_{b}$, $\epsilon_{e}$, $n_0$, $\theta_c$, $\theta_v$, and $p$) using the Nested Sampling library \sw{MultiNest}, implemented via \sw{PyMultiNest} \citep{2014A&A...564A.125B}, with 2000 live points. The best-fit model (Figure~\ref{fig:modelled_lc}) reproduces both the plateau and the sharp decay, with the achromatic break across all bands indicating a geometric jet break from an off-axis structured jet. Observed flux densities are shown as markers, median model light curves as dashed lines, and $3\sigma$ uncertainties as shaded bands. The priors and best-fit values are listed in Table~\ref{tab:mcmc_post}, with posterior distributions in Figure~\ref{fig:corner_plot} of the Appendix \ref{cornerplot}.

\begin{deluxetable*}{ccccc}[ht]
\tablewidth{0pt}
\tablecaption{Summary of the priors and posteriors for the model parameters obtained from \multi fitting for off-axis structured jet model. \label{tab:mcmc_post}}
\tablehead{
\colhead{Parameter} & \colhead{Unit} & \colhead{Prior Type} & \colhead{Parameter Bound} & \colhead{Posterior Value}
}
\startdata
$\log_{10}(E_\mathrm{K,iso})$ & erg & uniform & [51, 56] & $54.17 \pm 0.12$ \\
$\log_{10}(\epsilon_b)$ & \nodata & uniform & [-5, -1] & $-2.06 \pm 0.48$ \\
$\log_{10}(\epsilon_e)$ & \nodata & uniform & [-3, -0.5] & $-0.64^{+0.10}_{-0.15}$ \\
$\log_{10}(n_0)$ & cm$^{-3}$ & uniform & [-4, 1] & $-1.86 \pm 0.75$ \\
$\theta_c$ & rad & log-uniform & [$10^{-4}$, 0.2] & $0.012 \pm 0.003$ \\
$\theta_v$ & rad & log-uniform & [$10^{-4}$, 0.2] & $0.032 \pm 0.008$ \\
$p$ & \nodata & uniform & [2.001, 2.8] & $2.04 \pm 0.01$ \\
$\chi$ & \nodata & fixed & 1 & 1 \\
\enddata
\tablecomments{The posterior values are presented with their uncertainties, and parameter bounds are listed separately for clarity where applicable.}
\end{deluxetable*}


\giteam The posterior distributions for \thisgrb\ indicate a highly energetic jet with $E_\mathrm{K, iso} = (1.5 \pm 1.4) \times 10^{54}\ \mathrm{erg}$, consistent with the long-lived plateau observed for \thisgrb. The circum-burst medium density is constrained to $n_0 = 0.01~\mathrm{cm}^{-3}$, consistent with expectations for short GRBs \citep{2015ApJ...815..102F, 2020MNRAS.495.4782O}. The microphysical parameters are $\epsilon_e = 0.23^{+0.05}_{-0.07}$, $\epsilon_b = 0.008^{+0.016}_{-0.006}$, and $p = 2.04 \pm 0.02$. For the structured jet, we obtain a narrow core angle of $\theta_c \approx 0.69^{\circ}$ and a viewing angle of $\theta_v \approx 1.83^{\circ}$. This implies that 75\% of total the energy is concentrated within the narrow jet ($\simeq \theta_c$) and it declines at higher viewing angles. Within the obtained $\theta_v$, fraction of total energy within the jet is 98.5\%.


\giteam Figure~\ref{fig:model_spectra} shows the spectral energy distributions (SEDs) from the off-axis structured jet model at t=0.17~d, t=0.96~d, t=1.53~d, and t=4.04~d. At t=0.17~d, we include the temporally extrapolated spectrum from the light-curve fit, while at later epochs the observed optical data is shown at t=1.53~d and radio measurements at t=4.04~d. Across all epochs, the optical SEDs follow a consistent spectral slope $\beta$, demonstrating achromatic evolution in this regime. This agreement between the model and the sparse multi-wavelength data highlights the robustness of the structured jet interpretation.

\giteam Next, we examine the evolution of the synchrotron break frequencies, $\nu_c$ and $\nu_m$, during the observed light curve. Since the jet is being observed significantly off-axis, we cannot use simple analytic estimates for the evolution of these frequencies with time. Instead, we derive the temporal behavior of these frequencies from the best-fit model at different epochs. As seen in the bottom panel of Figure~\ref{fig:modelled_lc}, we find that the cooling frequency, $\nu_c$, remains above all the observed bands throughout the duration of the observations. The characteristic frequency, $\nu_m$, crosses the optical band at very early times before observations commence, and subsequently passes through the radio band at $\sim 4$~d after the GRB trigger.

\giteam Based on the evolution of the synchrotron break frequencies, most of the optical afterglow is observed in the adiabatic cooling regime ($\nu_m < \nu_{\mathrm{optical}} < \nu_c$) barring a few early data points. In this regime, the closure relation for a jet expanding into a uniform ISM predicts a spectral slope of $\beta = (p-1)/2 = 0.52 \pm 0.01$, which is comparable to the measured spectral indices ($\beta_{\mathrm{o1}}$, $\beta_{\mathrm{o2}}$) derived in \S\ref{sec:afterglow_spectral}. In contrast, the radio band lies in the $\nu_{\mathrm{radio}} < \nu_m$ regime, where a positive spectrum with $F_{\nu} \propto \nu^{1/3}$ is expected. However, the spectral index $\beta_{\mathrm{radio}}$ derived in \S\ref{sec:afterglow_spectral} shows a negative slope: inconsistent with standard synchrotron theory, but consistent with jet break from an off-axis structured jet. Furthermore, the temporal decay index ($\alpha$) does not solely follow the standard closure relations of synchrotron emission but is also modulated by the viewing geometry of an off-axis structured jet.



The radiative efficiency of a GRB quantifies the fraction of the total energy budget emitted as prompt $\gamma$-rays \citep{2004ApJ...613..477L}. If we directly calculate this value for \thisgrb, we get a very low number:  $\eta \sim 0.3\%$, rather than the expected 10--20\% range expected for the fireball model with internal shocks \citep{1999ApJ...523L.113K, 2009ApJ...707.1623M, Wang_2015}. The reason for this apparent discrepancy is that the efficiency definition is to be applied for an on-axis observer, where the observed emission is dominated by material along the line of sight. This shows the importance of ascertaining whether the observer is within the jet core before interpreting the apparent jet efficiencies. To consistently compare the jet kinetic energy during the afterglow phase ($E_{\mathrm{K,iso}}$) with the prompt emission in the efficiency calculation, the observed (line of sight) prompt energy $E_{\mathrm{iso}}$ must be corrected to its on-axis (core) value. In this case, the Doppler factor does not play a role \citep{2024MNRAS.533.1629O}, and the prompt $\gamma$ ray energy can be approximated from the jet’s angular energy profile (Equation~\ref{eq:powerlaw_jet}), assuming no angular dependence of the $\gamma$-ray efficiency. 

\section{Discussion}\label{sec:discussion}

\subsection{Structure of the jet}
\giteam The light curve of \thisgrb\ shows several distinctive features compared to a typical GRBs: (1) a pronounced long plateau in the optical light curve from the first detection at 10.5~min to the break time, $t_b = 0.96 \pm 0.02$~days; (2) a steep postbreak decay with $\alpha = 3.29 \pm 0.18$; (3) an achromatic break in both X-ray and optical bands; (4) negative spectral decay indices ($\beta$) before and after the break across all bands. 

A nominal model consisting of a top-hat jet expanding into an ISM environment cannot create long plateaus. In most cases, the light curves show a rise, peak, and a decline. An observer located just beyond the jet core at say $\theta_v/\theta_c = 1.1$ may observe short plateaus \citep[see for instance][]{2008MNRAS.387..497P}, but extending the duration to $\sim10^5$~s would make the plateau unrealistically bright.
Top-hat jet afterglows do show achromatic jet breaks, but these typically occur after a relatively shallow decay phase. In contrast, the light curve of \thisgrb\ exhibits a plateau followed by a very steep decline, which is inconsistent with the jet-break scenario.

\giteam On the other hand, a power-law structured jet observed off-axis can naturally produce both a late-time peak and a long-lasting plateau \citep{2008MNRAS.387..497P}. For a general structured jet, the isotropic equivalent energy of the blast wave depends on the angle from the jet axis; expressed as $E(\theta) = 4\pi dE/d\Omega$. The launch mechanism initially sets the angular structure of the jet and can subsequently be modified by interaction with the external medium. In the case of short GRBs associated with compact object mergers, the ejecta along the polar direction is relatively sparse. Hence, the jet structure remains largely unchanged from its original form \citep{2020ApJ...896..166R}. 




For an off-axis structured jet model, the early rise and plateau phases are highly sensitive to the jet structure, and can be used to infer jet geometry \citep{2024ApJ...975..131R, 2021ApJ...909..114N}. In contrast, the post-peak declining phase of the light curve provides limited constraints on the jet geometry, since it closely resembles the evolution of an on-axis afterglow. Moreover, the ratio $\theta_v / \theta_c$ modulates the morphology of the light curve before peak. If this ratio is close to unity, the light curve shows a decay; for larger ratios, a rising behavior is observed. The peak in the light curve is determined by both jet's core angle and the viewing-to-core angle ratio. For \thisgrb, we infer a narrow jet with $\theta_c \sim 0.7^{\circ}$ and an intermediate ratio of $\theta_v / \theta_c \approx 2.73$, which naturally explains the observed long plateau followed by a steeper decay, consistent with an early-peaking light curve.

\subsection{Comparison with GW170817}

\begin{figure*}[th]
    \centering
    \begin{subfigure}[t]{0.48\textwidth}
        \includegraphics[width=\linewidth]{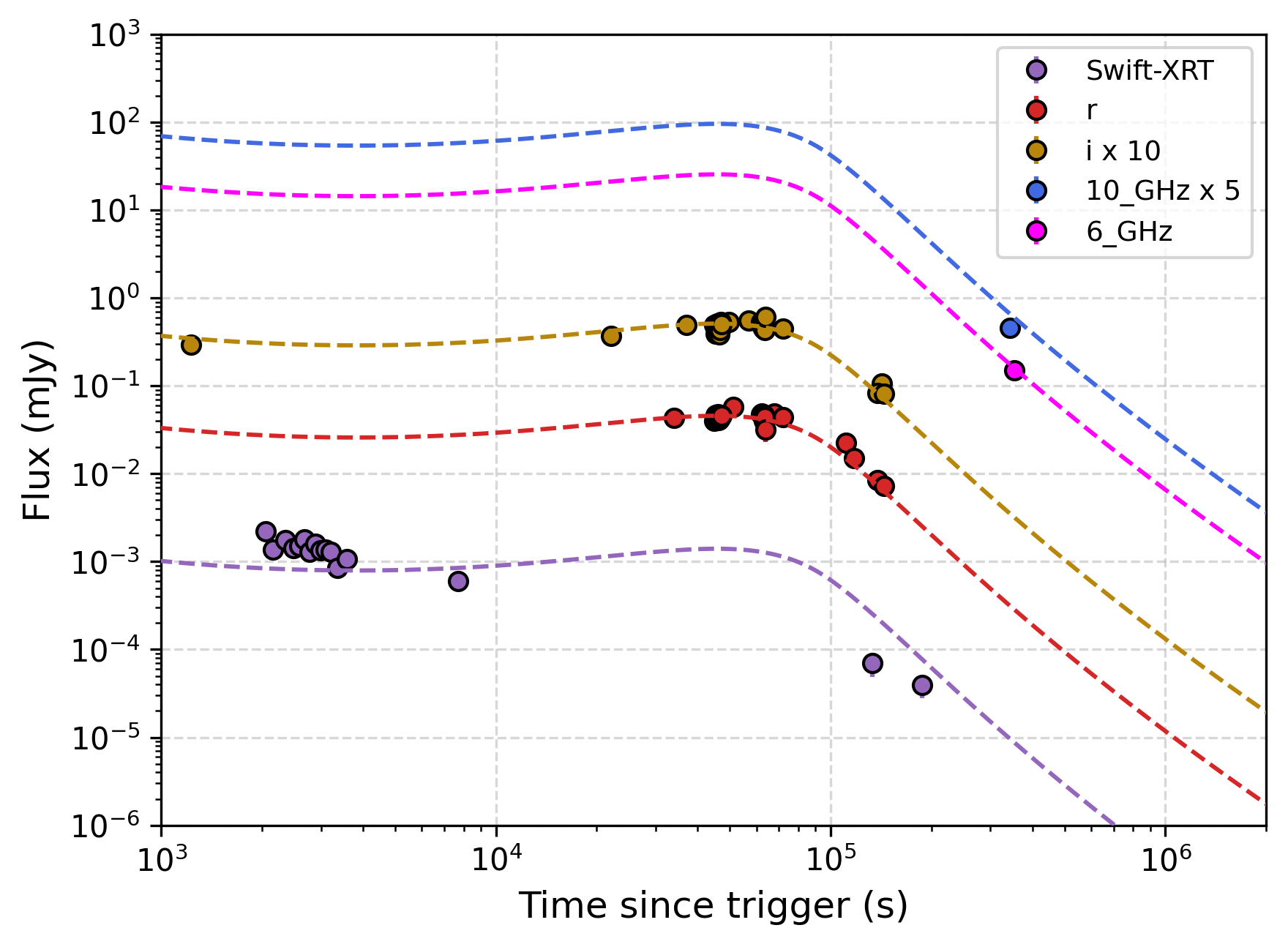}
        \caption{Model originally fit to GW170817 \citep{2024ApJ...975..131R} but recomputed using the core angle and viewing-angle ratio inferred for \thisgrb\ ($\theta_v - \theta_c \approx 1.1^{\circ}$, $\theta_v/\theta_c \approx 2.7$). The modeled flux is scaled up to match observed afterglow of \thisgrb.}
        \label{fig:model-1}
    \end{subfigure}%
    \hfill
    \begin{subfigure}[t]{0.48\textwidth}
        \includegraphics[width=\linewidth]{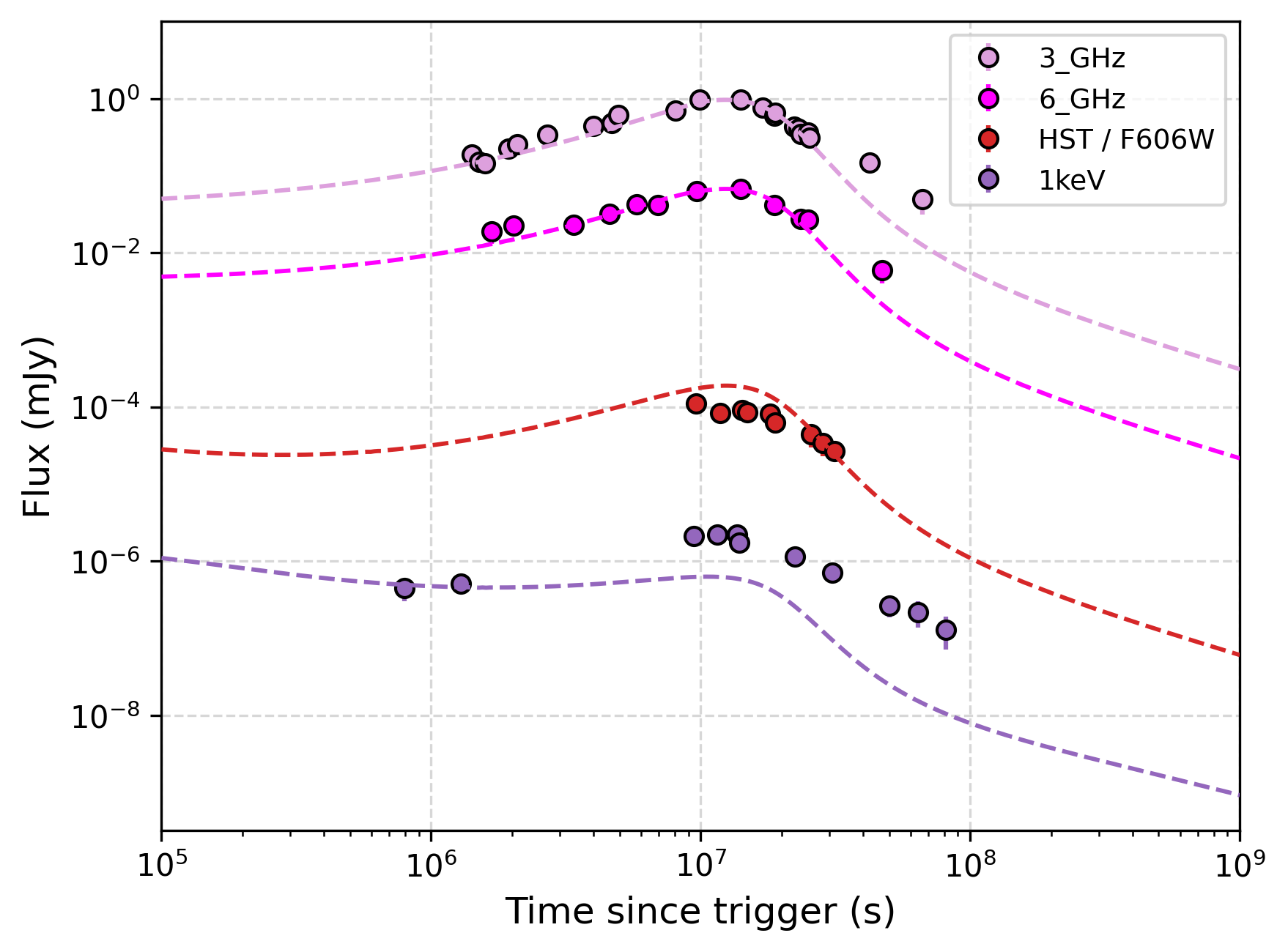}
        \caption{Model from our fit to \thisgrb\ recomputed using the core angle and viewing-angle ratio inferred for GW170817 ($\theta_v - \theta_c \approx 14^{\circ}$, $\theta_v/\theta_c \approx 4.5$). The afterglow data of GW170817 event is collected from \citep{2021ApJ...922..154M}.}
        \label{fig:model-2}
    \end{subfigure}
    \caption{Comparison of structured-jet geometries between \thisgrb\ and the GW170817 afterglow.}
    \label{fig:gw170817-model}
\end{figure*}


\giteam 
The GRB afterglow of the GW170817 had, in addition to a detailed light curve \citep[][and references therein]{2021ApJ...922..154M}, a VLBI measurement of its image at super-luminal motion across the sky \citep{2018Natur.561..355M,2019Sci...363..968G,2022Natur.610..273M}. The image motion confirmed that the afterglow is generated by an off-axis jet and enabled a tight measurement of the viewing angle,   $\theta_v \approx 19^{\circ}$ with an estimated error of a few degrees, and jet core angle ($\theta_c \approx 1.5^{\circ}-4^{\circ}$) at the time of the peak, about 150 days after the merger \citep{2022Natur.610..273M,Govreen2023}. A detailed numerical modeling of the jet expansion has shown that the initial jet opening angle (before spreading) was in the range $\sim 0.5^{\circ}-4^{\circ}$ and that the observations are best explained by a power-law structured jet with $s \approx  3-4$ \citep{Govreen2024}. Due to degeneracies between the model parameters, the density into which the jet propagated is not well constrained, and it was probably around $10^{-3}~\mathrm{cm}^{-3}$, with an uncertainty of at least an order of magnitude \citep[see for instance][]{2019MNRAS.489.1919T}. Also, the initial jet isotropic equivalent energy in the core is not well constrained, and it most likely was about one or two orders of magnitude lower than the value that we infer for \thisgrb . In our discussion, we adopt the values of \( \theta_c = 4\degr \) and \( \theta_v / \theta_c = 4.5 \) for GW170817.


For an off-axis GRB, time at which the afterglow peaks depends on the viewing geometry, and the ratio of the isotropic equivalent kinetic energy to the circum-burst density, but is only weakly sensitive to the core angle \citep{Govreen2024}. In particular, it is proportional to \( (\theta_v - \theta_c)^2 \). Since this value is about 14\degr\ for GW170817 but only $\sim 1.1\degr$ for \thisgrb, the light curve peak shifts from $\sim162$~d \citep{2019MNRAS.489.1919T} to $\sim 1$~d.


We undertake a direct comparison between GW170817 and \thisgrb. First, we take the model parameters of GW170817 from \citet{2024ApJ...975..131R}, and change \( \theta_c \) and \( \theta_v \) values to those of \thisgrb. We then apply a scale factor to the model, and resultant curves are shown in Figure~\ref{fig:model-1}, along with observed data. We see a good agreement in the two. Due to the low of \( \theta_v - \theta_c \) the light curve peaks much earlier, while the low value of \( \theta_c \) causes the early light curve to be a plateau rather than a rise, as discussed in \citet{Govreen2024}. Next, we take model parameters of \thisgrb\ but change \( \theta_c \) to 4\degr\ and \( \theta_v \) to 18\degr. The resultant light curves, again scaled by an overall factor, are shown in Figure~\ref{fig:model-2}. Despite other parameters being fit to \thisgrb\ data, the model shows reasonable correspondence with observed values for GW170817.


 We therefore conclude that both \thisgrb\ and GW170817A can be consistently described within the off-axis structured jet framework, with their contrasting light curve evolution being dominated by differences in jet and viewing geometry. In particular, the narrower jet core and intermediate viewing-angle ratio of \thisgrb\ explain its earlier peak and extended plateau, while the broader jet and larger ratio of GW170817 result in the much later peak. This comparison highlights the importance of long-term follow-up of short GRBs, since structured jets with larger $\theta_v - \theta_c$ values may remain undetected until very late times.

\begin{acknowledgments}
We thank all members of the GROWTH collaboration for helping with observations and data processing.

The GROWTH India Telescope \citep[GIT; ][]{2022AJ....164...90K} is a 70-cm telescope with a 0.7-degree field of view, set up by the Indian Institute of Astrophysics (IIA) and the Indian Institute of Technology Bombay (IITB) with funding from DST-SERB and IUSSTF. It is located at the Indian Astronomical Observatory (Hanle), operated by IIA. We acknowledge funding by the IITB alumni batch of 1994, which partially supports the operations of the telescope. Telescope technical details are available at \url{https://sites.google.com/view/growthindia/}.

This work is partially based on data obtained with the 2m Himalayan Chandra Telescope of the Indian Astronomical Observatory (IAO) under the proposal 2025-02-P10 (PI: D. Eappachen). We thank the staff of IAO, Hanle, and CREST, Hosakote, that made these observations possible. The facilities at IAO and CREST are operated by the Indian Institute of Astrophysics, Bangalore. We thank the staff of the GMRT that made these observations possible. GMRT is run by the National Centre for Radio Astrophysics of the Tata Institute of Fundamental Research.

Some of the data presented herein were obtained at Keck Observatory, which is a private 501(c)3 non-profit organization operated as a scientific partnership among the California Institute of Technology, the University of California, and the National Aeronautics and Space Administration. The Observatory was made possible by the generous financial support of the W. M. Keck Foundation. The authors wish to recognize and acknowledge the very significant cultural role and reverence that the summit of Maunakea has always had within the Native Hawaiian community. We are most fortunate to have the opportunity to conduct observations from this mountain. 

MWC acknowledges support from the National Science Foundation with grant numbers PHY-2117997, PHY-2308862 and PHY-2409481.

NS  is supported by a fellowship from the Kavli foundation.

AP acknowledges support by the National Science Foundation AST Grant No. 2308193. This project used data obtained with the Dark Energy Camera (DECam), which was constructed by the Dark Energy Survey (DES) collaboration. Based on observations at Cerro Tololo Inter-American Observatory, NSF's NOIRLab (NOIRLab Prop. ID 2025A-729671, PI: Palmese), which is managed by the Association of Universities for Research in Astronomy (AURA) under a cooperative agreement with the National Science Foundation.

MK and VB acknowledge the support from the VAIBHAV Fellowship of the Department of Science and Technology of the Government of India.

VB and the IIT Bombay team sincerely thank BDP UGL Global Logistics Pvt. Ltd. for their generous CSR support towards our computational needs.

We thank Anirudh Salgundi and Arjun Ghosh for help with the Chandra proposal. We thank Ashwin Devaraj for useful discussions.

GCA thanks the Indian National Science Academy (INSA) for support under their Senior Scientist Programme.

The scientific results reported in this article are based on observations made by the Chandra X-ray Observatory. This research has made use of software provided by the Chandra X-ray Center (CXC) in the application packages CIAO.

S.~A. is supported by an LSST-DA Catalyst Fellowship, funded through the support of Grant 62192 from the John Templeton Foundation to LSST-DA. S.~A. also gratefully acknowledges support from Stanford University, the United States Department of Energy, and Fred Kavli and The Kavli Foundation.

The Andreoni Transient Astronomy Lab is supported by the National Science Foundation award AST 2505775, NASA grant 24-ADAP24-0159, and the Discovery Alliance Catalyst Fellowship Mentors award 2025-62192-CM-19.

\end{acknowledgments}

\begin{contribution}

All authors contributed equally to this collaborative work.


\end{contribution}

%
\facilities{\emph{Konus}-Wind, \emph{Swift}(XRT), \emph{Chandra}, GIT:0.7m, HCT:2m, Keck:10m, Blanco:8m, Fraunhofer:2m, Palomar 200-inch, Palomar 60-inch, uGMRT.}

\software{Astropy \citep{2013A&A...558A..33A,2018AJ....156..123A,2022ApJ...935..167A}, Source Extractor \citep{1996A&AS..117..393B}, Astro-SCRAPPY \citep{2019ascl.soft07032M}, \sw{solve-field} astrometry engine \citep{2010AJ....139.1782L}, \sw{PSFEx} \citep{2013ascl.soft01001B}}, \jetsimpy \citep{2024ApJS..273...17W}, \texttt{ciao-4.15} \citep{ciao}, \afterglowpy \citep{2020ApJ...896..166R, 2024ApJ...975..131R}, \sw{PyMultiNest} \citep{2014A&A...564A.125B}, \texttt{CASA} \citep{2022PASP..134k4501C}

\appendix 

\section{Afterglow data}\label{appendixData}

In this section we list and describe the photometry collected for the analysis of \thisgrb, along with the reduction and calibration methods. 

\subsection{GIT}\label{GIT}
\giteam We used the GIT located at the Indian Astronomical Observatory (IAO), Hanle-Ladakh, to acquire data of the optical afterglow of \thisgrb\ \citep{2025GCN.40962....1M}. GIT is a 0.7-meter wide-field, fully robotic telescope specifically designed for the study of transient astrophysical events \citep{2022AJ....164...90K}. The afterglow was observed in the Sloan $r{^\prime}$ and $i{^\prime}$ filters. Data were downloaded and processed in real time using the GIT data reduction pipeline. All images were pre-processed by subtracting bias \& flat-fielding followed by cosmic-ray removal via Astro-SCRAPPY \citep{2019ascl.soft07032M} package. Astrometry was performed on the resulting images using the offline \sw{solve-field} astrometry engine \citep{2010AJ....139.1782L}. The sources were detected using \sw{SExtractor} \citep{1996A&AS..117..393B} and crossed matched with the PanSTARRS DR1 catalog \citep{2016arXiv161205560C} through \texttt{vizier} to obtain the zero point in the images. Finally, the pipeline performed point spread function (PSF) photometry using \sw{PSFEx} \citep{2013ascl.soft01001B} to generate the PSF of the image and obtain the \thisgrb\ afterglow magnitudes.

\subsection{\emph{Swift}-XRT}
\giteam \emph{Swift}-XRT began observing \thisgrb\ approximately 34~min after the trigger and continued with multiple epochs up to 2.17~days. For the light curve analysis, we used the publicly available results from the UK Swift Science Data Centre\footnote{\url{https://www.swift.ac.uk/xrt_live_cat/00021535/}}. We adopted the unabsorbed flux and photon index ($\Gamma$) reported in the 0.3--10~keV energy range. Assuming a powerlaw spectrum of the form $F_{\nu} \propto \nu^{-\beta}$, where the spectral index $\beta = \Gamma -1$ and $\Gamma$ is the photon index. Using the unabsorbed band flux, we normalized the power-law spectrum over the $0.3-10$~keV range and then evaluated the corresponding flux density at 1~keV and 10~keV.

\subsection{\emph{Chandra}}
\giteam We triggered \emph{Chandra} through \emph{Chandra} DDT (proposal number 26409057, PI Pathak) to observe \thisgrb. The source was observed for a single epoch at $\sim 18.10$ days from the trigger for an exposure of $19.82$~ks. We reprocessed the data to get a new level-2 data through \texttt{ciao-4.15}. Since the source was not detected, we followed \texttt{srcflux} method to determine the model-dependent upper limit, and obtained flux  $ \le 3.41 \times 10^{-15}~\textrm{erg cm}^{-2}\textrm{s}^{-1}$ in the energy range of 0.5-7.0~keV. With same method discussed in previous section, we calculated flux density of $5.34 \times 10^{-8}$~mJy at 10~keV.

\subsection{HCT}
\giteam We observed the field of \thisgrb\ using the Himalayan Faint Object Spectrograph Camera mounted on the 2m Himalayan Chandra Telescope (HCT) at the Indian Astronomical Observatory (IAO), Hanle, India. Observations were carried out in $J$ and $r{^\prime}$ bands, beginning at 2025-07-05T21:16:47.94 UT for the $J$ band and at 2025-07-07T17:12:23.27 UT for the $r{^\prime}$ band. A total exposure time of 1170~s in the $J$ band and 3600~s in the $r^\prime$ band was obtained. Standard data reduction and photometric analysis were performed using Astro-SCRAPPY, \sw{SExtractor}, the offline astrometry.net algorithm, and \sw{PSFEx}, as in the case of GIT. The magnitudes are calibrated against PanSTARRS DR1 for $r^\prime$ band and against 2MASS catalog \citep{2006AJ....131.1163S} for the $J$ band. The derived upper limits are listed in Table~\ref{tab:afterglow_obs_table}.

\subsection{Keck}

We observed \thisgrb\ with MOSFIRE mounted on the 10~m Keck I telescope (PI Kasliwal, PROGID: C348), and acquired J-band imaging of the afterglow. The observations started at 08:55 UT on July 9, 2025, and consisted of five sets of box-9 dithered images with 11~s exposures and three coadds each. We used standard reduction methods to coadd the images and used the 2MASS catalog to calibrate our photometry. We detect a source close to the 3$\sigma$ limit of our observations at J = 24.4 $\pm$ 0.2 mag (AB). 

Additionally, we observed \thisgrb\ with the Low-Resolution Imaging Spectrometer (LRIS; Oke et al. 1995) on the Keck I telescope. Observations started 10:20 UT on July 24, 2025, and consisted of a series of twenty 30~s exposures using the 680 dichroic simultaneously the V and I filters. We do not find a source at the position of the afterglow.

\subsection{Blanco}
We observed with Blanco with four epochs from July 5th to July 8th. On the first epoch data was taken with the $ugriz$ filters, on the 6th and 7th data was taken with $riz$, on the 8th data was taken solely in $r$.

\subsection{Fraunhofer Telescope}

We observed with the 3kk instrument mounted on the Fraunhofer Telescope at Wendelstein Observatory (FTW) using the $r$, $i$, and $J$ bands \citep{lang2016wendelstein}. We acquired 4 epochs of data on 2025-07-04 20:50:45, 2025-07-05 01:28:10, 2025-07-05 22:34:31, and 2025-07-06 00:23:53 UTC \citep{2025GCN.40974....1B}. Each night two epochs were acquired, one at the beginning of the night and one at the end. The first 3 epochs were taken with $10 \times 180$~s exposures, and the last epoch was taken with $30 \times 180$~s exposures. We calibrate the J-band observations against the 2MASS catalog and the $r$ and $i$ band against the Pan-STARRS1 catalog. We detect the afterglow in all epochs.


\subsection{Palomar 200-inch}
We use the the Wide-field Infrared Camera (WIRC; \citealt{2003SPIE.4841..451W}) on the Palomar 200-inch telescope using the near-infrared J and Ks bands. We acquired 3 epochs of WIRC data on July 04--06, 2025. The observations consisted on 3 sets of box-9 dithered images of 45~s and 1 coadd for J-band and 3~s and 10 coadds for Ks-band. We followed standard reduction techniques and calibrated against 2MASS. We detect the afterglow in the images in the first and second epoch, while not on the third epoch.

\subsection{Palomar 60-inch}
We acquired images with the Spectral Energy Distribution Machine (SEDM; \citealt{SEDM}) on on July 04, 2025 and July 05, 2025. The first epoch was automatically scheduled through our program responding to Einstein Probe events. Our second epoch was in response to the afterglow detection. We follow standard reduction methods and calibrate against Pan-STARRS \citep{2016A&A...593A..68F}.

\subsection{uGMRT}
We observed the field of \thisgrb\ with the wideband receiver backend of the upgraded Giant Metrewave radio Telescope (uGMRT) in two frequency bands - band 4 (central frequency 750~MHz, bandwidth 400~MHz) and band 5 (central frequency 1260~MHz, bandwidth 400~MHz) on 16 July 2025 and 17 July 2025 respectively (48\_059, PI: Eappachen). The raw data were downloaded in the \texttt{FITS} format and converted to the \texttt{CASA} \citep{2022PASP..134k4501C} measurement set format. Then the data was calibrated and imaged using the automated continuum imaging pipeline \texttt{CASA-CAPTURE} \citep{2021ExA....51...95K}. Eight rounds of self calibration were done within each pipeline run. Both the band 4 and band 5 observations in both epochs did not yield detections, and the upper limit values listed in Table~\ref{tab:radio-data} are the $3\times$RMS value in a large circle (of radius $\sim20\times$ resolution at the respective band) centered at the location of the \thisgrb, in the residual image.

\subsection{Other Public Data}

We collected the photometry circulated through GCN on this target (see Table \ref{tab:afterglow_obs_table}), which spans from $g$-band to J-band. 

\startlongtable
\begin{deluxetable*}{cCcccccc}
\tablecaption{Multi-wavelength afterglow observations of \thisgrb\ from optical bands. The table includes the time since the burst ($T - T_0$) in seconds, filter or band, central frequency (in Hz), measured magnitude (in AB system), upper limits, galactic extinction corrected magnitude (in AB system), and observing instrument along with references for each data point.} \label{tab:afterglow_obs_table}
\tabletypesize{\scriptsize}
\tablehead{
\colhead{Time - $T_0$} & \colhead{Filter} & \colhead{Frequency} & \colhead{Mag} & \colhead{Lim Mag} & \colhead{Corr Mag} & \colhead{Instrument} & \colhead{Ref.} \\
\colhead{(sec)} & \colhead{} & \colhead{($\times 10^{14}$ Hz)} & \colhead{AB} & \colhead{AB} & \colhead{AB} & \colhead{} & \colhead{}
}
\startdata
37044 & i & 3.931700 & 19.90$\pm$0.10 & -- & 19.67$\pm$0.10 & JinShan & \cite{2025GCN.40965....1L} \\
1224 & i & 3.931700 & 20.46$\pm$0.06 & -- & 20.23$\pm$0.06 & COLIBRI & \cite{2025GCN.40942....1S} \\
2568 & z & 3.282160 & 20.12$\pm$0.05 & -- & 19.95$\pm$0.05 & ESO-VLT-FORS2 & \cite{2025GCN.40945....1M} \\
16848 & VT\_B & 6.050000 & 20.40$\pm$0.20 & -- & 19.92$\pm$0.20 & SVOM/VT & \cite{2025GCN.40960....1X} \\
16848 & VT\_R & 3.810000 & 20.40$\pm$0.20 & -- & 20.19$\pm$0.20 & SVOM/VT & \cite{2025GCN.40960....1X} \\
22053 & i & 3.931700 & 20.20$\pm$0.03 & -- & 19.97$\pm$0.03 & Panstarrs & \cite{2025GCN.40958....1G} \\
22053 & z & 3.282160 & 20.01$\pm$0.07 & -- & 19.84$\pm$0.07 & Panstarrs & \cite{2025GCN.40958....1G} \\
34101 & r & 4.811310 & 20.12$\pm$0.08 & -- & 19.82$\pm$0.08 & GIT & This work \\
49648 & i & 3.931700 & 19.81$\pm$0.09 & -- & 19.58$\pm$0.09 & GIT & This work \\
51115 & r & 4.811310 & 19.80$\pm$0.09 & -- & 19.50$\pm$0.09 & GIT & This work \\
57708 & i & 3.931700 & 19.78$\pm$0.05 & -- & 19.55$\pm$0.05 & NOT & \cite{2025GCN.40971....1M} \\
63108 & r & 4.811310 & 20.20$\pm$0.10 & -- & 19.90$\pm$0.10 & FTW-3KK & \cite{2025GCN.40974....1B} \\
63108 & i & 3.931700 & 19.90$\pm$0.10 & -- & 19.67$\pm$0.10 & FTW-3KK & \cite{2025GCN.40974....1B} \\
63108 & J & 2.400000 & 19.40$\pm$0.20 & -- & 19.31$\pm$0.20 & FTW-3KK & \cite{2025GCN.40974....1B} \\
67680 & r & 4.811310 & 20.00$\pm$0.20 & -- & 19.70$\pm$0.20 & ESO-VLT-UT3 & \cite{2025GCN.40966....1A} \\
72000 & g & 6.284960 & 20.20$\pm$0.10 & -- & 19.76$\pm$0.10 & GSP, LCO & \cite{2025GCN.40975....1L} \\
72000 & r & 4.811310 & 20.10$\pm$0.10 & -- & 19.80$\pm$0.10 & GSP, LCO & \cite{2025GCN.40975....1L} \\
72000 & i & 3.931700 & 20.00$\pm$0.10 & -- & 19.77$\pm$0.10 & GSP, LCO & \cite{2025GCN.40975....1L} \\
72000 & J & 2.400000 & 19.50$\pm$0.10 & -- & 19.41$\pm$0.10 & ESO-VLT-UT4-HAWK-I & \cite{2025GCN.40970....1Y} \\
111111 & r & 4.811310 & 20.83$\pm$0.05 & -- & 20.53$\pm$0.05 & GIT & This work \\
118102 & R & 4.686720 & 21.06$\pm$0.05 & -- & 20.77$\pm$0.05 & AZT-33IK, Mondy & \cite{2025GCN.41024....1V} \\
142344 & i & 3.931700 & 21.56$\pm$0.14 & -- & 21.33$\pm$0.14 & NOT & \cite{2025GCN.40993....1A} \\
460800 & J & 2.400000 & 24.40$\pm$0.20 & -- & 24.31$\pm$0.20 & Keck-MOSFIRE & This work \\
45258 & r & 4.811310 & 20.04$\pm$0.12 & -- & 19.74$\pm$0.12 & FTW-3KK & This work \\
45553 & r & 4.811310 & 20.14$\pm$0.07 & -- & 19.84$\pm$0.07 & FTW-3KK & This work \\
45764 & r & 4.811310 & 20.04$\pm$0.06 & -- & 19.74$\pm$0.06 & FTW-3KK & This work \\
45976 & r & 4.811310 & 20.02$\pm$0.06 & -- & 19.72$\pm$0.06 & FTW-3KK & This work \\
46188 & r & 4.811310 & 20.19$\pm$0.08 & -- & 19.88$\pm$0.08 & FTW-3KK & This work \\
46399 & r & 4.811310 & 20.15$\pm$0.08 & -- & 19.85$\pm$0.08 & FTW-3KK & This work \\
46611 & r & 4.811310 & 20.11$\pm$0.07 & -- & 19.81$\pm$0.07 & FTW-3KK & This work \\
46823 & r & 4.811310 & 20.06$\pm$0.06 & -- & 19.76$\pm$0.06 & FTW-3KK & This work \\
47034 & r & 4.811310 & 20.03$\pm$0.06 & -- & 19.73$\pm$0.06 & FTW-3KK & This work \\
47246 & r & 4.811310 & 20.04$\pm$0.06 & -- & 19.74$\pm$0.06 & FTW-3KK & This work \\
61902 & r & 4.811310 & 20.02$\pm$0.03 & -- & 19.71$\pm$0.03 & FTW-3KK  & This work \\
62114 & r & 4.811310 & 20.03$\pm$0.05 & -- & 19.72$\pm$0.05 & FTW-3KK & This work \\
62326 & r & 4.811310 & 20.00$\pm$0.03 & -- & 19.70$\pm$0.03 & FTW-3KK & This work \\
62537 & r & 4.811310 & 20.04$\pm$0.04 & -- & 19.74$\pm$0.04 & FTW-3KK & This work \\
62749 & r & 4.811310 & 20.11$\pm$0.05 & -- & 19.80$\pm$0.05 & FTW-3KK & This work \\
62961 & r & 4.811310 & 20.09$\pm$0.05 & -- & 19.79$\pm$0.05 & FTW-3KK & This work \\
63172 & r & 4.811310 & 20.03$\pm$0.06 & -- & 19.73$\pm$0.06 & FTW-3KK & This work \\
63384 & r & 4.811310 & 20.09$\pm$0.05 & -- & 19.79$\pm$0.05 & FTW-3KK & This work \\
63596 & r & 4.811310 & 20.09$\pm$0.07 & -- & 19.79$\pm$0.07 & FTW-3KK & This work \\
63808 & r & 4.811310 & 20.45$\pm$0.27 & -- & 20.15$\pm$0.27 & FTW-3KK & This work \\
45258 & i & 3.931700 & 20.15$\pm$0.18 & -- & 19.93$\pm$0.18 & FTW-3KK & This work \\
45553 & i & 3.931700 & 19.92$\pm$0.07 & -- & 19.69$\pm$0.07 & FTW-3KK & This work \\
45764 & i & 3.931700 & 19.85$\pm$0.07 & -- & 19.62$\pm$0.07 & FTW-3KK & This work \\
45976 & i & 3.931700 & 20.08$\pm$0.09 & -- & 19.86$\pm$0.09 & FTW-3KK & This work \\
46188 & i & 3.931700 & 19.96$\pm$0.08 & -- & 19.73$\pm$0.08 & FTW-3KK & This work \\
46399 & i & 3.931700 & 20.16$\pm$0.10 & -- & 19.94$\pm$0.10 & FTW-3KK & This work \\
46611 & i & 3.931700 & 19.90$\pm$0.08 & -- & 19.68$\pm$0.08 & FTW-3KK & This work \\
46823 & i & 3.931700 & 20.03$\pm$0.08 & -- & 19.81$\pm$0.08 & FTW-3KK & This work \\
47034 & i & 3.931700 & 19.81$\pm$0.06 & -- & 19.59$\pm$0.06 & FTW-3KK & This work \\
47246 & i & 3.931700 & 19.87$\pm$0.06 & -- & 19.65$\pm$0.06 & FTW-3KK & This work \\
61902 & i & 3.931700 & 19.82$\pm$0.04 & -- & 19.59$\pm$0.04 & FTW-3KK & This work \\
62114 & i & 3.931700 & 19.92$\pm$0.06 & -- & 19.70$\pm$0.06 & FTW-3KK & This work \\
62326 & i & 3.931700 & 19.89$\pm$0.04 & -- & 19.66$\pm$0.04 & FTW-3KK & This work \\
62537 & i & 3.931700 & 19.87$\pm$0.05 & -- & 19.64$\pm$0.05 & FTW-3KK & This work \\
62749 & i & 3.931700 & 19.99$\pm$0.06 & -- & 19.76$\pm$0.06 & FTW-3KK & This work \\
62961 & i & 3.931700 & 19.93$\pm$0.06 & -- & 19.70$\pm$0.06 & FTW-3KK & This work \\
63172 & i & 3.931700 & 19.98$\pm$0.09 & -- & 19.75$\pm$0.09 & FTW-3KK & This work \\
63384 & i & 3.931700 & 19.92$\pm$0.07 & -- & 19.69$\pm$0.07 & FTW-3KK & This work \\
63596 & i & 3.931700 & 20.04$\pm$0.10 & -- & 19.81$\pm$0.10 & FTW-3KK & This work \\
63808 & i & 3.931700 & 19.67$\pm$0.24 & -- & 19.44$\pm$0.24 & FTW-3KK & This work \\
45272 & J & 2.400000 & 19.84$\pm$0.34 & -- & 19.75$\pm$0.34 & FTW-3KK & This work \\
45567 & J & 2.400000 & 19.42$\pm$0.14 & -- & 19.33$\pm$0.14 & FTW-3KK & This work \\
45779 & J & 2.400000 & 19.81$\pm$0.21 & -- & 19.72$\pm$0.21 & FTW-3KK & This work \\
45990 & J & 2.400000 & 19.59$\pm$0.17 & -- & 19.50$\pm$0.17 & FTW-3KK & This work \\
46202 & J & 2.400000 & 19.90$\pm$0.26 & -- & 19.81$\pm$0.26 & FTW-3KK & This work \\
46414 & J & 2.400000 & 20.11$\pm$0.31 & -- & 20.02$\pm$0.31 & FTW-3KK & This work \\
46625 & J & 2.400000 & 19.91$\pm$0.24 & -- & 19.82$\pm$0.24 & FTW-3KK & This work \\
46837 & J & 2.400000 & 19.48$\pm$0.15 & -- & 19.39$\pm$0.15 & FTW-3KK & This work \\
47048 & J & 2.400000 & 19.37$\pm$0.13 & -- & 19.28$\pm$0.13 & FTW-3KK & This work \\
47260 & J & 2.400000 & 19.41$\pm$0.14 & -- & 19.32$\pm$0.14 & FTW-3KK & This work \\
61918 & J & 2.400000 & 19.53$\pm$0.08 & -- & 19.43$\pm$0.08 & FTW-3KK & This work \\
62128 & J & 2.400000 & 19.68$\pm$0.13 & -- & 19.58$\pm$0.13 & FTW-3KK & This work \\
62340 & J & 2.400000 & 19.52$\pm$0.08 & -- & 19.43$\pm$0.08 & FTW-3KK & This work \\
62552 & J & 2.400000 & 19.44$\pm$0.07 & -- & 19.35$\pm$0.07 & FTW-3KK & This work \\
62763 & J & 2.400000 & 19.51$\pm$0.08 & -- & 19.42$\pm$0.08 & FTW-3KK & This work \\
62976 & J & 2.400000 & 19.44$\pm$0.07 & -- & 19.35$\pm$0.07 & FTW-3KK & This work \\
63187 & J & 2.400000 & 19.57$\pm$0.08 & -- & 19.48$\pm$0.08 & FTW-3KK & This work \\
63399 & J & 2.400000 & 19.58$\pm$0.09 & -- & 19.49$\pm$0.09 & FTW-3KK & This work \\
63610 & J & 2.400000 & 19.65$\pm$0.09 & -- & 19.56$\pm$0.09 & FTW-3KK & This work \\
63822 & J & 2.400000 & 19.72$\pm$0.34 & -- & 19.62$\pm$0.34 & FTW-3KK & This work \\
137884 & r & 4.811310 & 21.89$\pm$0.07 & -- & 21.58$\pm$0.07 & FTW-3KK & This work \\
144446 & r & 4.811310 & 22.06$\pm$0.05 & -- & 21.76$\pm$0.05 & FTW-3KK & This work \\
137884 & i & 3.931700 & 21.84$\pm$0.11 & -- & 21.61$\pm$0.11 & FTW-3KK & This work \\
144446 & i & 3.931700 & 21.85$\pm$0.06 & -- & 21.62$\pm$0.06 & FTW-3KK & This work \\
137898 & J & 2.400000 & 21.65$\pm$0.27 & -- & 21.56$\pm$0.27 & FTW-3KK & This work \\
144461 & J & 2.400000 & 21.19$\pm$0.10 & -- & 21.10$\pm$0.10 & FTW-3KK & This work \\
133056 & J & 2.400000 & -- & 18.90 & -- & HCT & This work \\
291168 & r & 4.811310 & -- & 21.80 & -- & HCT & This work \\
\enddata
\end{deluxetable*}

\begin{deluxetable*}{cccccc}
\tablecaption{Log of X-ray observations of the X-ray afterglow of
GRB 250704B taken using Swift-XRT and Chandra. \label{tab:xray-data}}
\tablewidth{0pt}
\tablehead{
\colhead{$T_\mathrm{start}-T_{0}$} & \colhead{$T_\mathrm{stop}-T_{0}$} & \colhead{Flux} & \colhead{Photon Index} & \colhead{Flux$_{10~\mathrm{keV}}$} & \colhead{Flux$_{1~\mathrm{keV}}$} \\
\colhead{(s)} & \colhead{(s)} & \colhead{($10^{-11}$ $\mathrm{erg}~\mathrm{cm}^{-2}~\mathrm{s}^{-1}$
 )} & \colhead{} & \colhead{($10^{-4}$ mJy)} & \colhead{($10^{-3}$ mJy)}
}
\startdata
$1963$ & $2109$ & $2.43 \pm 0.54$ & $1.65^{+0.43}_{-0.31}$ & $4.97 \pm 1.10$ & $2.22 \pm 0.49$ \\ 
$2109$ & $2214$ & $1.41 \pm 0.32$ & $1.72^{+0.31}_{-0.22}$ & $2.60 \pm 0.59$ & $1.37 \pm 0.31$ \\ 
$2214$ & $2435$ & $1.64 \pm 0.37$ & $1.84^{+0.28}_{-0.20}$ & $2.52 \pm 0.57$ & $1.75 \pm 0.39$ \\ 
$2435$ & $2535$ & $1.26 \pm 0.28$ & $1.93^{+0.33}_{-0.23}$ & $1.68 \pm 0.37$ & $1.43 \pm 0.32$ \\ 
$2535$ & $2633$ & $1.27 \pm 0.28$ & $2.03^{+0.24}_{-0.18}$ & $1.41 \pm 0.31$ & $1.52 \pm 0.34$ \\ 
$2633$ & $2718$ & $1.44 \pm 0.32$ & $2.12^{+0.31}_{-0.23}$ & $1.37 \pm 0.30$ & $1.80 \pm 0.40$ \\ 
$2718$ & $2834$ & $1.00 \pm 0.23$ & $2.19^{+0.24}_{-0.18}$ & $0.83 \pm 0.19$ & $1.28 \pm 0.29$ \\ 
$2834$ & $2926$ & $1.22 \pm 0.28$ & $2.26^{+0.24}_{-0.18}$ & $0.88 \pm 0.20$ & $1.61 \pm 0.36$ \\ 
$2926$ & $3044$ & $1.00 \pm 0.22$ & $2.31^{+0.31}_{-0.23}$ & $0.64 \pm 0.14$ & $1.33 \pm 0.30$ \\ 
$3044$ & $3159$ & $1.05 \pm 0.23$ & $2.25^{+0.23}_{-0.18}$ & $0.77 \pm 0.17$ & $1.38 \pm 0.30$ \\ 
$3159$ & $3272$ & $1.01 \pm 0.23$ & $2.18^{+0.20}_{-0.15}$ & $0.84 \pm 0.19$ & $1.30 \pm 0.30$ \\ 
$3272$ & $3445$ & $0.68 \pm 0.15$ & $2.11^{+0.24}_{-0.18}$ & $0.66 \pm 0.15$ & $0.84 \pm 0.19$ \\ 
$3445$ & $3681$ & $0.87 \pm 0.15$ & $2.08^{+0.26}_{-0.20}$ & $0.89 \pm 0.15$ & $1.07 \pm 0.18$ \\ 
$7571$ & $7832$ & $0.50 \pm 0.11$ & $2.04^{+0.21}_{-0.16}$ & $0.54 \pm 0.12$ & $0.60 \pm 0.13$ \\ 
$92754$ & $172449$ & $0.063 \pm 0.019$ & $1.90^{+0.54}_{-0.39}$ & $0.088 \pm 0.026$ & $0.070 \pm 0.021$ \\ 
$177117$ & $193406$ & $0.036 \pm 0.010$ & $1.88^{+0.59}_{-0.43}$ & $0.052 \pm 0.015$ & $0.039 \pm 0.011$ \\
$1553549^{\star}$ & $1573369$ & $<0.003$ & - & $<0.0053$ & - \\
\enddata
\tablecomments{Flux values (col.~3) are obtained in the 0.3--10 keV band. Flux$_{10~\mathrm{keV}}$ and Flux$_{1~\mathrm{keV}}$ are the flux densities calculated at 10 keV and 1 keV, respectively.\\ $\star$ observation taken with Chandra.}
\end{deluxetable*}

\begin{deluxetable*}{ccccccc}
\tablecaption{Log of radio data for the radio afterglow of GRB 250704B taken using VLA, MeerKAT, and uGMRT. \label{tab:radio-data}}
\tablehead{
\colhead{$T-T_{0}$ (s)} & \colhead{Instrument} & \colhead{Energy-band} & \colhead{Flux (mJy)}  & \colhead{Flux upper lim (mJy)} & \colhead{Reference}
}
\startdata
354240    & VLA     & 6 GHz   & $0.150 \pm 0.005$ & -     & \cite{2025GCN.41038....1S} \\
343872    & VLA     & 10 GHz  & $0.092 \pm 0.007$ & -     & \cite{2025GCN.41046....1R} \\
492480    & MeerKAT     & 1.3 GHz & $0.070 \pm 0.005$ & -     & \cite{2025GCN.41060....1S} \\
1171802.9 & uGMRT & 1.3 GHz &- & 0.050    & This work \\
1171620.9 & uGMRT & 0.65 GHz & -  & 0.114 & This work \\
\enddata
\end{deluxetable*}

\section{Corner Plot of Structured Jet Model} \label{cornerplot}
\begin{figure*}[ht]
    \centering
    \includegraphics[width=\textwidth]{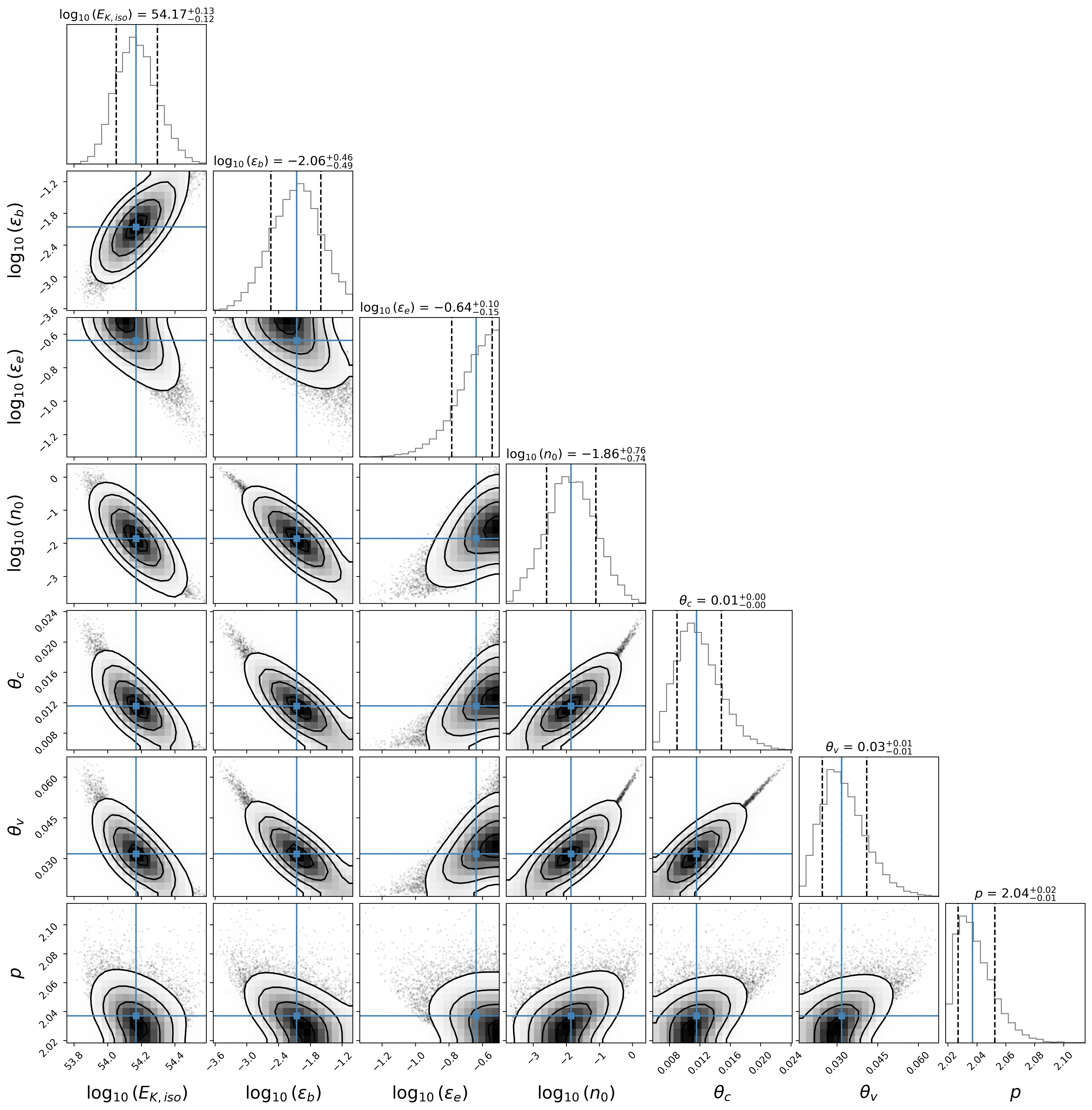}
    \caption{Posterior distribution of physical parameters for model fitted using \jetsimpy\ with structured jet interacting with ISM medium and \multi. The model fit for the $\log_{10}(E_{\mathrm{K,iso}})$, $\log_{10}(\epsilon_{b})$, $\log_{10}(\epsilon_{e})$, $\log_{10}(n_0)$, $\theta_c$, $\theta_v$ and $p$ parameters. The histogram shows the 16 per cent, 50 per cent, and 84 per cent percentiles of the probability distribution.} 
    \label{fig:corner_plot}
\end{figure*}

For completeness, Figure~\ref{fig:corner_plot} shows the corner plot of the posterior distributions of the parameters obtained from our off-axis structured jet modeling (see \S\ref{sec:jetsimpy}).

\section{Energy injection modeling} \label{appendixEnergyInjection}
In this section we explore models including energy injection.

The presence of long plateaus in GRB afterglows is often attributed to continued energy injection by some central engine, for instance by a magnetar \citep{2001ApJ...552L..35Z, 2008MNRAS.385.1455M, Rowlinson_2013}.
To test this scenario, we used the \afterglowpy\ package, where the energy injection is parameterized as:
\begin{equation}
L(t) = L_0 \left(\frac{t}{t_0}\right)^{-q},
\end{equation}
with $L_0$ is luminosity of energy injection, $t_0$ fixed to $1$~ks by default, and $q$ power law index of energy injection \citep{2020ApJ...896..166R, 2024ApJ...975..131R}. An additional parameter, $t_s$, specifies the time in the source frame at which the injection ceases. 

In our modeling, we adopted $q=0$, which corresponds to a nearly constant luminosity injection from the central engine over the time scale $t_s$, followed by a sudden termination when the central engine collapses to a black hole. Such constant injection naturally explains the shallow decay or plateau phase commonly observed in many GRB afterglows \citep{2001ApJ...552L..35Z, 1998A&A...333L..87D}. For the dynamics of the jet, we assumed a relativistic top-hat jet propagating in a uniform ISM, with free parameters $E_{\mathrm{K,iso}}$, $\epsilon_b$, $n_0$, $\theta_c$, $\theta_v$, and $p$.

\begin{figure}[ht]
    \centering
    \includegraphics[width=\columnwidth]{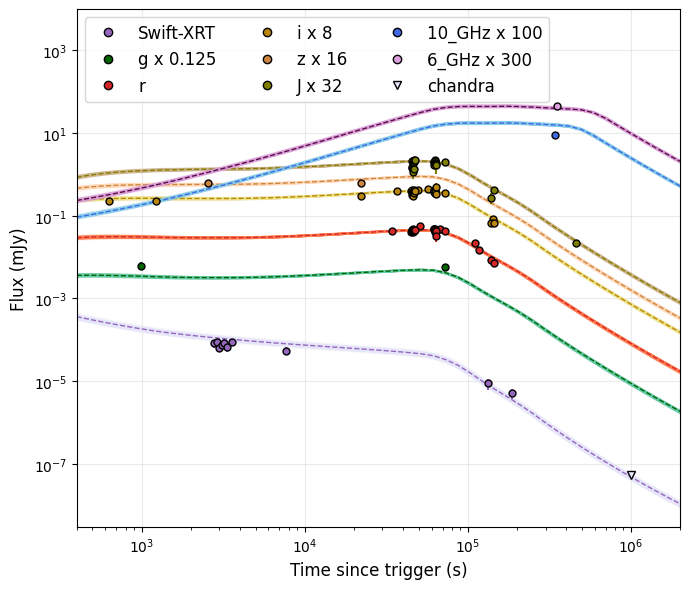}
    \caption{We modeled the multi-band afterglow light curves of \thisgrb\ using \afterglowpy, assuming a relativistic structured jet with a top-hat profile propagating into a uniform-density interstellar medium (ISM), and incorporating a constant energy-injection rate. Dotted lines show the best-fit light curves, and shaded regions mark the $3\sigma$ uncertainties.}
    \label{fig:modelled_lc_magnetar}
\end{figure}

Using the same sampling method described in the previous section, we obtained the posterior distributions summarized in Table~\ref{tab:mcmc_post_magnetar}. From the best-fit values, we infer an isotropic kinetic energy of $E_{\rm K,iso} \sim 10^{51}$ erg, implying a radiative efficiency exceeding 80\%. The model favors a jet with a core angle of $2.4^{\circ}$, viewed almost on-axis. We find a characteristic time of $t_s \sim 1$~ks in source frame, suggesting that up to this epoch the central engine remains active, continuously injecting energy into the jet.

The inferred circum-burst medium density is $n \sim 0.1\ \rm cm^{-3}$, about an order of magnitude larger than the range typically expected for magnetar-powered afterglows, $10^{-3} - 10^{-2}\ \rm cm^{-3}$ \citep{Rowlinson_2013}, where lower densities are generally more favorable for efficient energy injection. The microphysical parameters are $\epsilon_e = 0.74 \pm 0.07$, $\epsilon_b \sim 0.003$, and $p = 2.06 \pm 0.01$. The unusually high value of $\epsilon_e$ suggests a strong inverse Compton cooling \citep{2001ApJ...548..787S}, which is not included in our present model. On the other hand, if we fix $\epsilon_e$ to the typical value of 0.1, the fit quality degrades significantly and requires an unrealistically large external density, again inconsistent with the magnetar scenario. 

Overall, we find that while the constant energy injection model is capable of reproducing plateau features in GRB afterglows, fitting the afterglow of \thisgrb\ within this framework requires implausible values for physical parameters such as $\epsilon_e$ or $n_0$. In contrast, the off-axis structured jet model does not suffer from these shortcomings and is therefore preferred for explaining the afterglow of \thisgrb.

\begin{deluxetable*}{ccccc}[ht]
\tablewidth{0pt}
\tablecaption{Summary of the priors and posteriors for the model parameters obtained from \multi\ fitting of energy injection model implemented using \afterglowpy. \label{tab:mcmc_post_magnetar}}
\tablehead{
\colhead{Parameter} & \colhead{Unit} & \colhead{Prior Type} & \colhead{Parameter Bound} & \colhead{Posterior Value}
}
\startdata
$\log_{10}(E_\mathrm{K,iso})$ & erg & uniform & [48, 53] & $51.04^{+0.03}_{-0.03}$ \\
$\log_{10}(\epsilon_b)$ & --- & uniform & [-5, -1] & $-2.58^{+0.08}_{-0.06}$ \\
$\log_{10}(\epsilon_e)$ & --- & uniform & [-3, -0.1] & $-0.13^{+0.02}_{-0.04}$ \\
$\log_{10}(n_0)$ & $cm^{-3}$ & uniform & [-4, 1] & $-0.96^{+0.03}_{-0.04}$ \\
$\theta_c$ & rad & log-uniform & [$10^{-4}$, 0.2] & $0.044^{+0.001}_{-0.001}$ \\
$\theta_v$ & rad & log-uniform & [$10^{-4}$, 0.2] & $0.003^{+0.001}_{-0.001}$ \\
$p$ & --- & uniform & [2.001, 3] & $2.06^{+0.007}_{-0.004}$ \\  
$\log_{10}(L_{0})$ & erg~s$^{-1}$ & uniform & [43, 51] & $45.28^{+0.028}_{-0.023}$ \\
$\log_{10}(t_{s})$ & s & uniform & [2, 6] & $3.06^{+0.09}_{-0.10}$ \\
$\chi$ & --- & --- & --- & 1 \\
\enddata
\end{deluxetable*}



\bibliography{sample7}{}
\bibliographystyle{aasjournal}




\end{document}